\documentclass[10pt]{article}
\usepackage{a4}
\begin{document}
\begin{titlepage}
\title{On the Microscopic Perspective of Black Branes Thermodynamic Geometry}
\author{}
\date{ Stefano Bellucci  \thanks{\noindent bellucci@lnf.infn.it},
 Bhupendra Nath Tiwari \thanks{\noindent bntiwari.iitk@gmail.com}\\
\vspace{0.50 cm}
INFN-Laboratori Nazionali di Frascati\\
Via E. Fermi 40, 00044 Frascati, Italy.\\
\vspace{0.50 cm}
Department of Physics \\
Indian Institute of Technology-Kanpur,\\
Kanpur-208016, India.}
\vspace{0.50 cm}

\maketitle

\abstract{ In this article we study correspondence between the microscopic spectrum and
macroscopic properties of a class of extremal and non-extremal black branes and
outline an origin of the interactions among various microstates of a given black brane
configuration from the perspective of an intrinsic Riemannian geometry
arising from the coarse graining entropy over a large number of microstates.
We have analyzed the state-space geometry in the case of various extremal and non-extremal
black branes arising from the string theories, multi-centered black brane configurations,
small black holes with fractional branes, fuzzy rings in the set up of Mathur's fuzzballs
and subensemble theory, as well as that the black brane foams from the considerations
of bubbling black brane solutions in the M-theory.
We have further shown that there exists a clear mechanism on the black brane side
that describes the notion of associated interactions in the state-space or vice-versa.
We thus find that in all such cases there are no singularities in the state-space manifold
of these black brane configurations.
This observation is in turn consistent with the existing picture of corresponding
microscopic CFT data.}

\vspace{3cm}



\newpage
\end{titlepage}
\newpage
\begin{Large} \textbf{Contents:} \end{Large}\\
\begin{enumerate}

\item{Introduction.}
\item{Thermodynamic Geometry.}
\item{Black Holes in String Theory:}
\subitem{3.1 \ \ Extremal Black Holes.}
\subitem{3.2 \ \ Non-extremal Black Holes.}
\item{Multi-centered Black Branes: $D_6D_4D_2D_0$ Configuration.}
\item{Fractionation of Branes: Small Black Holes.}
\item{Mathur's Fuzzball Proposal and Subensemble Theory: Fuzzy Rings}
\subitem{6.1 \ \ The Fuzzball Proposal.}
\subitem{6.2 \ \ Subensemble Theory.}
\item{Bubbling Black Brane Solutions: Black Brane Foams:}
\subitem{7.1 \ \ A Toy Model: Single GH-center.}
\subitem{7.2 \ \ Black Brane Foams.}
\item{Discussion and Conclusion.}
\end{enumerate}

\section{Introduction}

In recent years, string theory has made significant progress
towards the understanding of the microstates for the extremal and
near extremal black branes. In particular, we can count the
microstates for certain black brane configurations that carry the
same charges and energy as the black brane with an entropy
$S_{micro} $ that equals to the Bekenstein-hawking entropy
$S_{BH}= \frac{A}{4G}$ of the corresponding black brane
\cite{9601029v2, 9602043v2}. The concept of the AdS/CFT
correspondence, see \cite{9711200v3, 0206126, 0212204} and
references therein, suggests that the $S_{micro}$ counts the
states of the branes in a field theory description dual to the
gravitational description. In fact, such conventional
understanding of the entropy is bases on coarse graining over a
large number of microstates and thus it turns out to be a crucial
ingredient in the realization of an equilibrium microstates
thermodynamic geometry.

It is known that the charged extremal black holes in $D=4,
\mathcal N=2 $ supergravity may be characterized by certain
electric and magnetic charges $q_J$ and $ p^I $ arising from usual
flux integrals of the field strength tensors and their
Poincar\'{e} duals. On the other hand, the scalar fields arising
from the compactification of either a string theory or M-theory
serve as the moduli which in fact parameterize the compact
internal manifold. The extremal charged black hole solutions are
BPS solitons, which interpolate between asymptotic infinity and
the near horizon geometry. The spherical symmetry in turn
determines this interpolation to be a radial evolution of the
scalar moduli which encodes the consequent changes in the
underlying internal compact manifold. Moreover, one has flat
Minkowskian manifold at the asymptotic infinity and thus the
asymptotic ADM mass for a given scalar moduli tending to certain
arbitrary values is described by the associated complex central
extension $Z_{\infty}$ of the $\mathcal N=2 $ supersymmetry
algebra, and in particular, it tuns out that $M(p,q,\phi^a)= \vert
Z_{\infty}\vert$, see for details \cite{9508072v3, 9602111v3}.

In such cases, the near horizon geometry of an extremal black hole
turns out to be an $AdS_2 \times S^2 $ manifold which describes
the concerned Bertotti-Robinson vacuum. The area of the horizon
$A$ and hence the macroscopic entropy is given as $ S_{macro}= \pi
\vert Z_{\infty} \vert^2 $. However, it turns out that the radial
variation of the moduli is described by a damped geodesic equation
which flows to an attractive fixed point at the horizon and thus
may solely be determined by the charges carried by the extremal
black hole. Such attractors may further be studied from the
perspective of criticality of black hole effective potential in $
\mathcal N=2, D=4 $ supergravities coupled to $n_{V}$ abelian
vector multiplets for an asymptotically flat extremal black hole
background which in turn is described by $(2n_{V}+2)$-dyonic
charges and the $n_{V}$-complex scalar fields which in turn
parameterize the $n_{V}$-dimensional special K\"{a}hler manifold,
see for detail \cite{0702019v1} and references therein. For
reviews on these subjects one can see for example
\cite{Bellucci:2007ds, Bellucci:2006zz}.

In the study of the attractor horizon geometries of the $D=4$
dyonic extremal black holes in certain supergravity theories, it
tuns out that such a hole have a non-vanishing Bekenstein-Hawking
entropy, see for example \cite{0805.1310}. In this article, the
authors have outlined such analysis from 1/2-BPS and non-BPS
(non-supersymmetric) attractors with non-vanishing central charge
in the context of the $\mathcal N=2,D=4$ ungauged supergravity
coupled to the $n_{V}$-number of abelian vector multiplets.
Further, they have demonstrated a complete classification of the
orbits in the symplectic representation of the classical U-duality
group, as well as that of the moduli spaces associated with the
non-BPS attractors in the context of symmetric special K\"ahler
geometries. Furthermore, the case of non-extremal black branes may
be considered similarly by adding the corresponding anti-branes to
the extremal black brane configurations, and thus the computation
of the associated brane entropy may either be performed in the
microscopic or that of the macroscopic considerations, see for
example \cite{9602043v2, 0412287}, which in fact shows a match
between the $ S_{micro}$ and $ S_{macro}$ for a set of given brane
charges and some total mass added to the extremal configuration.

On other hand, the role of multi-centered black holes in the context of $D$-brane systems
has recently been discussed in \cite{0705.2564v1, 0702146v2}.
In these developments the authors have considered a special class of multi-centered
black brane molecular configurations for which there is single centered black hole solution
whose degeneracy is predicted by an exact formula to be non-zero,
and thus showed how such states of $D$-branes may be represented as 2-centered configurations.
Here, we shall focus our attention on a special class of black brane configurations,
which may be characterized by the bound states of $D_6$, $D_4$, $D_2$ and $D_0$-branes.
However, it turns out that the concerned entropies for each of these two configurations
characterized by the charge vector considered in \cite{0705.2564v1, 0702146v2} are of different orders,
and hence their role in producing the correct contribution(s) to the degeneracy formulas.
This in turn implies that the nature of the correlations present in the state-space is not manifest.
Thus we intend to analyze an associated intrinsic Riemannian geometric model,
whose metric in the entropy representation may be defined as the negative Hessian matrix
of the entropy with respect to the extensive D-brane charges.
In turn, one can calculate the correlation functions and correlation volume directly
from the concerned entropy with such two centered black brane configurations
by using standard geometric techniques and thus may compare it with that
of the standard leading order single centered black brane configuration.
We find here that the associated geometric results are in a perfect agreement
with the fact that the multi-centered configurations are weakly interacting
than the most entropic minimal energy supergravity single centered configurations.

What follows next that we extend our study of the state-space geometry
of a given D-brane system into the perspective of the fractionation of branes
which in turn deals with the counting of certain chiral primaries.
For simplicity, we shall consider the example of two charge extremal small
black holes in type IIA string theory compactified on $ T^2 \times \mathcal M $,
where $ \mathcal M $ can be either $ K_3 $ or $ T^4 $, see for examples
\cite{9707203v1, 0507014v1, 0502157v4, 0505122v2}.
As the microscopic understanding of the entropy of the black holes in string theory
has been estimated in recent times at least for the case of two charge extremal black holes
\cite{0409148, 0411255, 0504005}.
There has thus been remarkable progress in our understanding of black holes having zero
horizon area in supergravity approximation but have non-zero statistical entropy from
independent string state counting arguments.
Such black hole space-time geometries are expected to develop non-zero horizon sizes
after including certain higher derivative corrections and in turn these black holes
are referred as small black holes.
The statistical entropy for a supersymmetric two-charge black hole coming from
counting the bound state degeneracy has been checked to agree with that of the
one calculated by using Wald's macroscopic entropy \cite{gr-qc/9307038, gr-qc/9502009, 9305016}
in higher-derivative supergravities \cite{0505122v2}.
The most studied interesting extremal black hole solutions are the small black holes,
which are the one made from a long heterotic string wrapping a circle carrying
winding charge $w$ and momentum charge $n$.
Here for simplicity, we consider our small black holes in the duality frame
which is described by the $D_0D_4$-brane configurations in the type-$IIA$
string theory compactified on $K_3 \times T^2$.

In particular, the microscopic understanding of this system is governed
by the superconformal quantum mechanics describing $D0$-branes in the
$AdS_2 \times S^2 \times CY_3$ attractor geometry of a black hole in
type IIA with $D4$-branes on the $CY_3$, see for details \cite{0412322}.
This quantum mechanics has a class of chiral primaries which may be identified
as the microstates of the black hole.
It was in turn shown in \cite{0412322} that the microscopic entropy for a small black hole
with vanishing triple intersection number could be reproduced from the leading order
asymptotic degeneracy of the associated chiral primaries.
Further from the microscopic perspective as shown in \cite{0409148},
the degeneracy in counting the microstates arises from the combinatorics
of the total $ N $ units of $ D_0 $-brane charge that splits into certain $ k $-small
clusters having $ n_i $ units of the $ D_0 $-brane charge
such that on each cluster the sum $ \sum_{i=1}^k n_i= N $ remains true
which in turn corresponds to the wrapped $ D_2 $-branes residing on either
of the $ 24 p $ bosonic chiral primary states.
It is thus instructive to ask: what are the possible interactions among the brane microstates
in a given k-cluster which are arising from the degeneracy of the two charged extremal
$D_0D_4$ small black holes?
In particular, we shall investigate such interactions from the perspective of
the brane fractionation in any finite number of arbitrary clusters and thus shed
light on the geometric perspective of the counting chiral primaries
associated with the two charge extremal small black holes.

On other hand, in order to understand how to count the
micro-states of a brane, Mathur propose the idea of brane
fractionation which leads to the fuzzball structure of the black
brane interior. Furthermore, the author suggests a way how one may
bypass the information paradox. For simplicity consider the case
of two charge extremal black hole, which is microscopically
characterized by certain string winding and momentum states. Then
the microscopic entropy comes from the number of different
possible ways in which the string carries the momentum as the
traveling waves. This is due to the fact that the associated
fundamental string in the heterotic picture does not carry any
longitudinal waves and thus all the vibrations must be transverse
vibrations. Thus the string carrying a momentum charge departs
from its central location $r=0$ and in turn gets spread over a
certain transverse region which makes a fuzzball.

Mathur further shows in \cite{0109154} that the different vibration states of
the string make different fuzzballs which do not have a singularity or a horizon.
However, if we draw a bounding sphere corresponding to the size of a typical fuzzball
then the area $A$ of the sphere satisfies $ {A\over G}\sim \sqrt{n_1n_p}\sim S_{micro} $,
see for more details \cite{0202072}.
In this way, the information contained in a black hole is distributed
throughout the interior of a horizon sized ball whose vicinity is not
the empty of information's but there exists radiation from an excited fuzzball
leaving the surface of such quantum fuzzball and thus carry out the information
about the state.
In fact, in any string theory construction, the bound states with certain momentum
cause the string to spread out over a horizon sized transverse region which
in turn generates the horizon sized fuzzball having sizable quantum fluctuations.

Moreover, it thus suggests that there exists a relation between the capped throat
structure of microstates geometries to that of the extremal black hole geometries
having zero entropy in the leading order gravity computations.
Note however that the actual microstates of the system in fuzzball conjecture do not have horizon,
but the boundary of underlying region where the microstates start differing from each other
satisfies a Bekenstein type relation with the entropy enclosed inside this boundary
whose state-space geometry we are going to investigate in this article.
Further we know that a black hole in general have a large entropy
which is a measure of the associated phase space volume and
thus the size of a typical fuzzball cannot be too small,
otherwise the phase space of the fuzzball
cannot account this large entropy of the black hole.

>From the view points of the subensemble theory, one may further carry out the investigations
of the state-space geometry for the rotating D1-D5 system by considering a family of
subensembles of states in which each boundary area of the fuzzball satisfies a Bekenstein
type relation with the entropy enclosed.
Recall further that the Mathur's fuzzball conjecture for an extremal black hole says that
the throat of the geometry ends in a finite quantum fuzz, in-contrast to the classical picture
with an infinite length horizon at the end of the throat, see for more details \cite{0706.3884v1}.

To explore further the relation between state-space interaction in a subensemble
with an enclosed entropy, we may restrict our states to a given subensemble.
The simplest case of this interaction may be described as follows.
Consider a string with given momentum charge then the corresponding string states
which can be either in high wave numbers with small transverse amplitude,
or in low wave numbers with large transverse amplitude.
Thus the entropy reaches an optimal value for some wave number and amplitude,
and such a typical state has infinite throat with certain horizon, which corresponds
to the classical picture of the black hole configuration.
However, if we restrict ourself to the subset of states which have
certain wave number higher than this optimal value and an amplitude
of transverse vibration smaller than the amplitude of the generic states.
Then the microscopic entropy of the states restricted to this smaller amplitude
is smaller than the standard microscopic entropy by a factor $M$,
which in turn is just the number of the subensembles of the considered ensemble.

In this case the string in the space-time description lives inside a smaller ball,
and thus the space-time geometry before capping off has a narrower and deeper throat.
It have been shown that the boundary of the fuzzball region in leading order computations
also decreases exactly by the same factor $M$, which further remains the same for the case
of with or without rotation.
These subensembles with smaller entropies have their fuzzball boundaries at deeper points
in the throat with smaller transverse areas and thus we find weaker state-space interactions
for two charge extremal systems with an angular momentum whose boundary are certain fuzzy regions
and in turn the associated area satisfies ${A\over G}\sim S_{micro}$ \cite{0706.3884v1}.
It is just for simplicity that we consider to deal with the state-space of two charged
rotating system.
But there are several evidences for three charge as well as four charge systems
that there exists analogous cap structures to that of the two charge systems
and thus one can easily describe the state-space manifolds associated with such
black branes in any subensemble under consideration.

As a final example of our goal, it is
one of the bubbling solutions whose state-space we offer to analyze
is the case of $D=11$ black brane solutions with given three charges having
required amount of supersymmetry.
It is known that in certain cases such solutions describe three charged BPS-black holes,
as considered in \cite{0209114,0401129,0408106,0408122}.
Recall in particular that in the case of the $T^6$ compactification of $M$-theory,
the $N^{th}$ Gibbon-Hawking charge far separated from the blob
is described by the associated 2-cycles $\Delta_{jN}$,
which must have a large flux compared to that of the fluxes on each
$\Delta_{ij}, \forall i,j < N$.
In such cases, these authors have shown that the flux parameters $ \lbrace k_i^I \rbrace $
may further be expressed in terms of the charges over the blob, whose quantized nature
arises from the gauge symmetry as the shift: $ k^I \rightarrow  k^I+  c^I V $ to be
half-integer.
It is to note that this shift symmetry may thus solely be determined by a function $V$
which in fact defines the nature of the Gibbon-Hawking base metric.
Further, it is known that the leading order charges in the limit of large N are given by
$Q_1 \simeq 4 N^2 k_0^2 k_0^3, Q_2 \simeq 4 N^2 k_0^1 k_0^3, Q_3 \simeq 4 N^2 k_0^1 k_0^2 $
and $ J_R= 16 N^3 k_0^1 k_0^2 k_0^3 $, which in fact turns out to be the case of a maximally
symmetric BMPV black hole with $J_R= 4 Q_1 Q_2 Q_3$, see for details \cite{0604110}.
Such relations however do not depend on the number of Gibbon-Hawking base points considered
and are further independent of the charges distributed on the Gibbon-Hawking base points.
See for instance \cite{0604110} for the case of the foaming of black ring solutions
with integer black ring dipole charge(s).

Thus one associates certain non-trivial entropy coming from the combinatorics of the
existing possibilities from the choices of $ \lbrace k_i^I \vert k_i^I> 0 \rbrace $
subject to a constraints for the summations over $k_i^I $
that the both worldvolume and supergravity (as well as that of the CFT) descriptions
yield the same relations between the parameters of the brane configurations.
Actually, it is the combinatorics of laying out such quantized fluxes on the topologically
non-trivial cycles $\Delta_{ij} $, which in turn may be referred as topological entropy
of the black brane.
Such an example appears in large charge limit with $J_1^2= J_2^2= Q_1Q_2Q_3$,
which in turn describes the associated leading order properties of the bubbled super-tubes,
see for details \cite{0505166}.
One of the natural question that we ask is:
how many bubbled black brane configurations exist for some given set of brane charges?
The other associated question that we further ask is:
what is the nature of the underlying state-space geometry?
In particular, whether there exist certain instabilities in a bubbled black brane configuration?
Note although that the partitioning of $k_i^1,k_i^2 $ and $k_i^3$ are not independent,
and thus a bubble configuration will collapse, if any of the flux parameter is taken to be zero.
So we need to count all the possible number of the ways of having non-zero partitions of the
$\lbrace k_i^I \rbrace$ over $N$ bubbles and finally sum up over the $N$ to obtain the
topological entropy of the black foam.
However, such effects turn out to be immaterial at the leading order consideration,
which is due to the fact that these effects contribute only to the subleading corrections
to the black foam entropy, see for details \cite{0608217}.

As a next step towards this objective,
it is instructive to explore the state-space geometry for the black brane foam configurations.
The example that we shall consider is the case of the M-theory compactified on $T^6$.
The simplest configuration turns out to have certain far separated unit Gibbon-Hawking charges
with total Gibbon-Hawking charge zero.
In particular, it is important to investigate whether the associated state-space metrics are non-degenerate
for such simplest configurations like the one above well separated Gibbon-Hawking charges
with vanishing total Gibbon-Hawking charge and thus we wish to establish the behavior of
the associated scalar curvature of the state-space manifold.
This construction would clearly elucidate the thermodynamical  issues of a black foam
and further provides a geometrical realization of the equilibrium thermodynamical
structures of the microstates of the foam configurations.
We demonstrate here that the state-space geometry of the black brane foam,
arising from the associated leading order topological entropy of the foam
is non-degenerate and everywhere regular for all well-defined three parameter
topological black brane foams.
Our results are thus in agreements with the fact that the black brane foams are
stable thermodynamic objects.
Furthermore, it may be argued that the nature of state-space of the given black brane foams
remains the same under the subleading corrections to the considered foam configurations.

We thus have a good understanding of the exact spectrum of a class of black branes
that there exits an asymptotic expansion of the degeneracy formula for large charges
which not only reproduces the entropy of the corresponding black branes to the leading order,
but also to the first few subleading orders as an expansion in inverse power of the charges \cite{0412287}.
Given this correspondence between microscopic spectrum and macroscopic entropies of a black brane,
a natural question to ask would be to understand an origin of the state-space interactions
from the degeneracy formula and it's correspondence to the black brane side?
In this paper, we indeed show from the perspective of an intrinsic Riemannian geometry
that there exists a clear mechanism on the black brane side that describes the associated
notion of the interactions among the microstates in the state-space of a brane solution
or vice-versa.

It is the thermodynamics of such black holes, black rings or in general black branes
that plays the central role in understanding certain properties of string theory or M-theory
\cite{9601029v2, 9508072v3, 9602111v3, 9607235, 9711053v1, 0408120v2, 0110260v2}.
In particular, the AdS/CFT correspondence involves equality of partition
functions of a string theory (or M-theory) on a space with asymptotics of
form $AdS_n \times K$, for some compact space $K$ for instance a Calabi-Yau,
and that of some holographic dual conformal field theory defined on the
conformal boundary of $AdS_n$
\cite{9905111v3, MOOREICTP, 0805.4216v1, 0805.0095v3, 0806.0053v2, 0707.3437v1}.
The present paper concentrates on the macroscopic-microscopic aspects of any
consistent black brane solutions and discusses the underlying state-space
geometry arising from the entropy obtained as certain bound states in the
theory of either D-branes or M-branes.
Our study is therefore well suited for the original applications of the
AdS/CFT correspondence and phase-transitions, if any, in the underlying
equilibrium state-space of a black brane configuration characterized
by certain electric-magnetic charges with or without rotations.

On other hand, the thermodynamic geometry is one of the main tool to study the
state-space structures of a class of black holes or black branes, see for instance
\cite{Weinhold, Ruppeiner, gr-qc/0512035v1, gr-qc/0601119v1, gr-qc/0304015v1,
0510139v3, 0606084v1, 0508023v2}.
It may further be seen that the thermodynamic geometry analyzes the equlibrium
microstates of such branes and in particular, the idea of the interactions or
phase transitions present in the state-space of such black branes may be
analyzed from the perspective of the macroscopic-microscopic duality.
In this article, we have analyzed various specific cases of the extremal
and non-extremal black holes, D-brane system like $D_6D_4D_2D_0$, fuzzy
black rings and bubbling black brane configurations for three charged
foamed black branes.
We find that such a thermodynamic system have regular state-space geometry
which turns out to be everywhere well-defined, and corresponds in general
to an interacting statistical systems, whenever there exists a non-zero
state-space scalar curvature.

The state-space geometry may further be shown to exhibit the associated microscopic
properties and thus one may analyze the geometric nature of the correlation
functions and correlation volume of the boundary dual conformal field theory
which in turn is in perfect accordance with the AdS/CFT correspondence.
This is because it is the components of the state-space metric tensor
that are related to the two point correlation functions of the boundary
conformal field theory, as the thermodynamic metric deals with Gaussian
fluctuations in the state-space of the brane configuration.
We have in turn explained how the two point correlation functions of certain
microstates characterizing a black brane behave in the state-space,
and in particular it is interesting to investigate whether they are regular
or singular functions in the state-space.
In the case of singular state-space scalar curvature, we have analyzed the
nature of the singularities present in the underlying state-space manifold.
For example, we have shown that there may exists certain critical points or
critical (hyper)-surfaces on which the state-space scalar curvature diverges.

In particular, the present article deals with the various implications arising
from the state-space geometry of the string theoretical extremal and non-extremal
black branes, multi-centered black brane configurations, fractionation of branes,
Mathur's subensemble theory and bubbling black brane solutions in M-theory.
This article has been organized into several sections.
The first section in turn motivated to study the state-space geometry obtained from
the Gaussian fluctuations of the entropy for given any consistent brane configuration
of either the string theory or M-theory, and in particular we have interesting
implications of the AdS/CFT correspondence.
In the section 2, we have introduced very briefly that what is the black
brane thermodynamic geometry bases on the large number of equilibrium microstates.
In the section 3, we analyze the state-space of extremal as well as
non-extremal black holes in string theory.
In the section 4, we focus our attention on the state-space geometry of
the multi-centered black brane configurations. Explicitly, we here discuss
the state-space geometry defined as an intrinsic Riemannian geometry obtained
from the entropy of $D_6D_4D_2D_0$ black brane configuration.

In the section 5, we investigate the implications arising from the fractionation
of $D$-branes on the state-space geometry of the two charged small black holes.
In the section 6, we would like to see the meaning of the state-space geometry
from the view-points of Mathur's fuzzballs and subensemble theory.
Here for simplicity we have demonstrated it explicitly for the example of
fuzzy rings.
In the section 7, we have extended our study of the state-space geometry to
the M-theory bubbling black brane solutions.
We have explained in this case that the state-space geometry obtained from
the entropy of a black brane foam is well-defined and pertains to an interacting
statistical system.
Finally, the section 8 contains certain concluding issues and discussions for
the state-space geometry of various black branes thus considered and their implications
to the boundary CFT data, and in particular Mathur's fuzzball proposal of a brane microstates.

\section{ Thermodynamic Geometry}

We begin by considering a brief review of an intrinsic Riemannian geometric model
whose covariant metric tensor is defined as negative Hessian of the entropy with
respect to the invariant electric-magnetic charges and angular momenta
carried by the black brane configuration.
In particular, for any n-dimensional state-space manifold $ M_n $,
we may define the components of the metric tensor of the so called Ruppenier geometry
\cite{Ruppeiner, 0510139v3, 0606084v1} to be
$ g_{ij}:=-\frac{\partial^2 S(\vec{x})}{\partial x^j \partial x^i} $,
where the vector $\vec{x}= (p^i,q_i, J_i) \in M_n $.
Explicitly for the case of two dimensional state-space geometry,
the components of the Ruppenier metric are given by, \\

$ g_{qq}=- \frac{\partial^2 S}{\partial q^2} $,

$ g_{qp}=- \frac{\partial^2 S}{{\partial q}{\partial p}} $, and

$ g_{pp}=- \frac{\partial^2 S}{\partial p^2} $.\\

Now we can  calculate the $\Gamma_{ijk}$, $R_{ijkl}$, $R_{ij}$
and $ R $ for above thermodynamic geometry $(M_2,g)$ and may
easily see that the scalar curvature is given by,\\

$ R= \frac{1}{2} (S_{qq}S_{pp}- S_{qp}^2)^{-2}
 (S_{pp}S_{qqq}S_{qpp}+ S_{qp}S_{qqp}S_{qpp}+ S_{qq}S_{qqp}S_{ppp}
-S_{qp}S_{qqq}S_{ppp}- S_{qq}S_{qpp}^2- S_{pp}S_{qqp}^2 )$.\\

Further, the relation between thermodynamic scalar curvature
and the Ruppenier curvature tensor for this state-space
geometry characterized by the equilibrium microstates of the
black hole configuration is given (see \cite{0606084v1} for details) by,
$ R=\frac{2}{\Vert g \Vert}R_{qpqp} $, which is quite usual for any
two dimensional intrinsic Riemannian manifold $(M_2(R),g)$.

Such a geometric formulation thus tacitly involves a statistical basis in terms of
a chosen ensemble or subensemble of the microstates of a considered black brane
configuration in the thermodynamic limit.
Here, it is worth to mention that the thermodynamic scalar curvature captures the
nature of the correlation volume present in the underlying statistical system.
This strongly suggests that in the context of the state-space manifold of a black brane
arising from either a consistent string theory or that of a $M$-theory solution as a closed systems,
the non-zero scalar curvature might provide useful information regarding the nature of the
interactions present between the microstates of the black brane thus considered.
With this general geometric introduction to the thermodynamic of a general brane solution,
let us now proceed to systematically analyze the state-space geometric structures
of such branes characterized by finite number of electric-magnetic charges with
or without angular momenta as a bound states of certain large number of $D$-branes
(or that of $M$-branes).

\section{Black Holes in String Theory}

In this section we shall study state-space geometry arising from
the entropy of black branes as a bound states of large number
D-branes having certain number of electric-magnetic charges,
angular momentum and may be mass in the case of non-extremal
black brane solutions.
It is well known that the extremal black branes do not Hawking radiate,
as since the Hawking temperature is proportional to the difference of the
inner and outer horizon radius of the brane configuration.

\subsection{Extremal Black Holes}
The simplest example of state-space geometry of a stringy
black hole that may be studied is as follows.
Consider at first to understand the state-space geometry arising
from an extremal black hole whose microstates are the winding and
the momentum modes of a string carrying $ n_1 $ number of winding
and $ n_p $ number of momentum, then the large charge microscopic
entropy turns out to be \cite{9309145v2, 9405117v1, 9411187v3, 9504147v2}:
$ S_{micro}= 2\sqrt{2n_1 n_p} $.

Macroscopically, this two charged black hole entropy may be
computed by considering $D_4$ and $D_0$ branes, with certain
compactification to obtain a $ M_{3,1} $ space-time. Of course in
string theory there are higher order corrections like $ R^2$ or $
R^4$-corrections for instance to usual Einstein action, and thus
these corrections makes non-zero horizon area, as it is being
stretched by such higher derivative corrections. These
computations of the macroscopic entropy are usually easy to solve
with spherically symmetric ansatz for the non-compact directions
\cite{0409148v2}. The microscopic entropy may be counted by the
consideration of weakly interacting D-brane ensemble
\cite{9512078v1}, and one finds for $n_4$ number of $D_4$ branes
and $n_0$ number of $D_0$ branes that $ S_{micro}= 2 \pi \sqrt{n_0
n_4}= S_{macro} $. Now the state-space geometry associated with
the entropy of $D_0, D_4 $ system may easily be seen to be
ill-defined. This is just from the determinant of the Hessian
matrix of the entropy with respect to all the extensive
thermodynamic variables which in this case are simply the charges
of $D_0, D_4 $ branes. This result holds for a constant entropy
curve for which the coordinates of the state-space lies on the
rectangular hyperbola: $ n_0 n_4= c^2 $. However, it is possible
to incorporate further higher derivative corrections to have well
defined Ruppenier like geometry where determinant of the Hessian
matrix of the entropy may become negative definite in certain
domain of the brane charges.

One of the important correction arrives due to the higher derivative
quartic terms in the Riemann tensor which may be encoded in a scalar
field $ \Upsilon $.
In this case, one finds that the space-time $ R^4 $-corrected macroscopic
entropy for the case of non-supersymmetric two charge small black holes
to be \cite{0606148v1, 0612225v2}: $ S(Q,P)= 2 \pi PQ(1+\frac{40C}{P^2}) $.
In order to take more closure look of the state-space geometry of
the equilibrium microstates of the two charged small black holes,
we shall now consider the implications of this $ R^4 $-corrected entropy
expression.
The Ruppeiner metric on the state-space may be easily read off from the
negative Hessian matrix of such a corrected entropy of the small black holes
to be,\\

$ g_{QQ}= 0 $,

$ g_{QP}= 2 \pi (-1+ \frac{40 C}{P^2}) $ and

$ g_{PP}= -\frac{160 \pi CQ}{P^3} $.\\

We observe here that the inclusion of higher derivative $ R^4 $-corrections
make the determinant of the metric tensor to be non-zero,
as one can simply see that it is just given by
$ \Vert g \Vert= -4 \pi^2 (-1+ \frac{40 C}{P^2})^2 $.
Moreover, the state-space of non-supersymmetric two charge small black holes
with quartic corrections corresponds to a non-interacting statistical system.
In particular, we find in this case that the scalar curvature is zero
and thus it certainly indicate thermodynamic stability and an absence
of phase transitions everywhere in the underlying state-space.
Here, we see that the constant entropy curve takes the form of a hyperbola:
$ \frac{\frac{1}{PQ}}{\frac{2 \pi}{k}}-\frac{\frac{1}{P^2}}{40C}=1 $
with coordinates $ \frac{1}{\sqrt{PQ}} $ and  $ \frac{1}{P} $, where the
constant $ k $ is some real number.

To have have more complicated black hole solutions we may add $ n_5 $ number
of $ D_5 $ branes, and then leading order black hole entropy obtained from
the Einstein action takes the form \cite{9601029v2}:
$ S_{micro}= 2 \pi \sqrt{n_1 n_5 n_p}= S_{macro} $.
In this of case again one can easily calculate the associated Ruppenier metric
and thus finds that the determinant of the metric tensor turns out to be a non-zero
quantity, as well as the fact that the associated scalar curvature is everywhere regular
which both scales as inverse of square root of the leading order $D_1 D_5$ branes entropy
with $ n_p $ number of Kluza-Klein momentum.
Here the constant entropy curve in the state-space is some higher dimensional
hyperbola: $ n_1 n_5 n_p= c^2 $.
Moreover, the similar results hold for four charge tree level extremal black holes
whose state-space geometry has the same form of metric tensor and scalar curvature
as that of the three charged extremal black hole configurations. See in
\cite{0801.4087v1,0806.3513v1} for certain interesting details of the thermodynamic
geometry applied to a large class of other black brane systems arising from the
string theory as well as that of the $M$-theory.

\subsection{Non-extremal Black Holes}

What follows next that we shall consider the state-space geometry arising
from the entropy of a non-extremal black hole, which one can simply make
just by adding the corresponding anti-branes to the extremal black brane
solutions.
The simplest example of such system is a string having large amount of
winding and $D_5$ brane charges $n_1, n_5$ with some extra energy,
which in the microscopic description creates the equal amount of momenta
running in the opposite direction of the $ S^1$.
In this case entropy has been calculated from both the microscopic and
macroscopic perspective \cite{9602043v2} which matches for some given
total mass and charges.
In particular, we have  $ S_{micro}= 2 \pi \sqrt{n_1 n_5} (\sqrt{n_p}+
\sqrt{ \overline{n_p}})= S_{macro} $.
The components of the Ruppeiner metric may now be computed easily as
the Hessian matrix of the corrected entropy expression to be,\\

$ g_{n_1 n_1}=  \frac{\pi}{2}\sqrt{\frac{n_5}{n_1}}(\sqrt{n_p}+
{\sqrt{\overline{n_p}}})$,

$ g_{n_1 n_5}= -\frac{\pi}{2\sqrt{n_1 n_5}} (\sqrt{n_p}+
{\sqrt{\overline{n_p}}})$,

$ g_{n_1 n_p}= -\frac{\pi}{2} \sqrt{\frac{n_5}{n_1 n_p}} $,

$ g_{n_1 \overline{n_p}}= -\frac{\pi}{2} \sqrt{\frac{n_5}{n_1 \overline{n_p}}} $,

$ g_{n_5 n_5}=  \frac{\pi}{2}\sqrt{\frac{n_1}{n_5}}(\sqrt{n_p}+
{\sqrt{\overline{n_p}}}) $,

$ g_{n_5 n_p}= -\frac{\pi}{2} \sqrt{\frac{n_1}{n_5 n_p}} $,

$ g_{n_5 \overline{n_p}}= -\frac{\pi}{2} \sqrt{\frac{n_1}{n_5 \overline{n_p}}} $,

$ g_{n_p n_p}= \frac{\pi}{2} \sqrt{\frac{n_1 n_5}{n_p^{3}}} $,

$ g_{n_p \overline{n_p}}= 0 $,

$ g_{\overline{n_p} \overline{n_p}}= \frac{\pi}{2} \sqrt{\frac{n_1 n_5}
{\overline{n_p}^{3}}} $.\\

The determinant of the metric tensor is
$ \Vert g \Vert= -\frac{\pi^4}{4 (n_p \overline{n_p})^{3/2}}(\sqrt{n_p}+
{\sqrt{\overline{n_p}}})^2$.
The determinant is non-zero for any set of given non-zero
brane-antibrane charges and as before we obtain a non-degenerate
state-space geometry of the equilibrium microstates of the four charged
non-extremal $ D_1 D_5 P $ black hole system.
The scalar curvature of underlying state-space may easily be computed
using the Ruppeiner metric technology defined as the negative Hessian matrix
of the entropy corrected by the non-extremal contributions.
In particular, we find that the state-space curvature scalar is given
to be
$ R(n_1, n_5, n_p, \overline{n_p})= \frac{9}{4 \pi \sqrt{n_1 n_5}}
(\sqrt{n_p}+ {\sqrt{\overline{n_p}}})^{-6}
[ n_p^{5/2}+ 10 n_p^{3/2} \overline{n_p})+ 5 n_p^{1/2}\overline{n_p}^{2}+
5 n_p^{2}\overline{n_p}^{1/2}+ 10 n_p \overline{n_p}^{3/2}+ \overline{n_p}^{5/2} ]$.
The curvature scalar is thus non-zero, except for the set of the roots of a
two variable polynomial for the two momenta $ (n_p, \overline{n_p}) $
running in the opposite directions of the $S^1$.
This polynomial of charges and anti-charges is simply defined as the
numerator of the state-space curvature scalar, that is just
$ f(n_p, \overline{n_p}):= n_p^{5/2}+ 10 n_p^{3/2} \overline{n_p})+
5 n_p^{1/2}\overline{n_p}^{2}+ 5 n_p^{2}\overline{n_p}^{1/2}+
10 n_p \overline{n_p}^{3/2}+ \overline{n_p}^{5/2} $.
Furthermore, the underlying state-space manifold is everywhere regular and
corresponds to a thermodynamic system whose statistical basis is
interacting except at the roots of the $ f(n_p, \overline{n_p}) $ and
thus there exists a hypersurface in the state-space of non-extremal
$ D_1 D_5 P $ black hole system on which it becomes non-interacting
statistical system.
Here, we can see that the constant entropy curve is certain non-standard
curve and it is just given by:
$ \frac{c^2}{n_1 n_5}= (\sqrt{n_p}+ \sqrt{ \overline{n_p}})^{2} $.

One can extrapolate the above entropy expression to a non-large
charge domain, where we are no longer close to a supersymmetric state.
The leading order entropy which includes all such special extremal and
near-extremal cases may be written as a function of charges
$ \lbrace n_i \rbrace $-anticharges $ \lbrace m_i \rbrace $
to be \cite{9603109v1}:
$ S(n_1,m_1,n_2,m_2,n_3,m_3):= 2 \pi (\sqrt{n_1} + \sqrt{m_1})
(\sqrt{ n_2 }+ \sqrt{m_2}) (\sqrt{n_3}+ \sqrt{m_3}) $.
It is again not difficult to explore the state-space geometry of
the equilibrium microstates of the 6 charge anti-charge non-extremal
black hole in $D=4$ arising from the entropy expression arising from
just the Einstein Action.
As stated earlier that the Ruppeiner metric on the state-space is given by
the negative Hessian matrix of the non-extremal entropy with respect to
the extensive variables.
These variables in this case are in turn the conserved charges-anticharges
carried by the non-extremal black hole. Explicitly, we find that
the components of covariant metric tensor under such non-large charge
domains are easily obtained to be,\\

$ g_{n_1 n_1}:= \frac{\pi}{n_1^{3/2}} ( \sqrt{n_2} + \sqrt{m_2} )
( \sqrt{n_3}+ \sqrt{m_3} ) $,

$ g_{n_1 m_1}:= 0 $,

$ g_{n_1 n_2}:= -\frac{\pi}{2 \sqrt{n_1 n_2}} ( \sqrt{n_3}+ \sqrt{m_3} )$,

$g_{n_1 m_2}:= -\frac{\pi}{2 \sqrt{n_1 m_2}} ( \sqrt{n_3}+ \sqrt{m_3} )$,

$g_{n_1 n_3}:= -\frac{\pi}{2 \sqrt{n_1 n_3}} ( \sqrt{n_2}+ \sqrt{m_2} )$,

$g_{n_1 m_3}:= -\frac{\pi}{2 \sqrt{n_1 m_3}} ( \sqrt{n_2}+ \sqrt{m_2} )$,

$g_{m_1 m_1}:= \frac{\pi}{m_1^{3/2}} ( \sqrt{n_2} + \sqrt{m_2} )
( \sqrt{n_3}+ \sqrt{m_3} )$,

$g_{m_1 n_2}:= -\frac{\pi}{2 \sqrt{m_1 n_2}} ( \sqrt{n_3}+ \sqrt{m_3} )$,

$g_{m_1 m_2}:= -\frac{\pi}{2 \sqrt{m_1 m_2}} ( \sqrt{n_3}+ \sqrt{m_3} )$,

$g_{m_1 n_3}:= -\frac{\pi}{2 \sqrt{m_1 n_3}} ( \sqrt{n_2}+ \sqrt{m_2} )$,

$ g_{m_1 m_3}:= -\frac{\pi}{2 \sqrt{m_1 m_3}} ( \sqrt{n_2}+ \sqrt{m_2} )$,

$g_{n_2 n_2}:= \frac{\pi}{n_2^{3/2}} ( \sqrt{n_1} + \sqrt{m_1} )
( \sqrt{n_3}+ \sqrt{m_3} )$,

$g_{n_2 m_2}:= 0$,

$g_{n_2 n_3}:= -\frac{\pi}{2 \sqrt{n_2 n_3}} ( \sqrt{n_1}+ \sqrt{m_1} )$,

$g_{n_2 m_3}:= -\frac{\pi}{2 \sqrt{n_2 m_3}} ( \sqrt{n_1}+ \sqrt{m_1} )$,

$g_{m_2 m_2}:= \frac{\pi}{m_2^{3/2}} ( \sqrt{n_1} + \sqrt{m_1} )
( \sqrt{n_3}+ \sqrt{m_3} )$,

$g_{m_2 n_3}:= -\frac{\pi}{2 \sqrt{m_2 n_3}} ( \sqrt{n_1}+ \sqrt{m_1} )$,

$g_{m_2 m_3}:= -\frac{\pi}{2 \sqrt{m_2 m_3}} ( \sqrt{n_1}+ \sqrt{m_1} )$,

$g_{n_3 n_3}:= \frac{\pi}{n_3^{3/2}} ( \sqrt{n_1} + \sqrt{m_1} )
( \sqrt{n_2}+ \sqrt{m_2} )$,

$g_{n_3 m_3}:= 0$,

$g_{m_3 m_3}:= \frac{\pi}{m_3^{3/2}} ( \sqrt{n_1} + \sqrt{m_1} )
( \sqrt{n_2}+ \sqrt{m_2} )$.\\

A straightforward computation yields the determinant of the metric tensor
to be $ \Vert g \Vert= -\frac{ \pi^6 }{16} (n_1m_1n_2m_2n_3m_3)^{-3/2}
( \sqrt{n_2} + \sqrt{m_2} )^2  (\sqrt{n_3} + \sqrt{m_3} )^3
(\sqrt{n_1} + \sqrt{m_1} )^3 ( n2 \sqrt{m1 n3} + n_2 \sqrt{m_1 m_3}
+ 2 \sqrt{n_2 m_1  m_2 n_3} + \sqrt{n_2 m_1  m_2 m_3}  +
m_2 \sqrt{m_1 n_3} +  m2 \sqrt{m_1 m_3} + n_2 \sqrt{n_1  n_3} +
n_2 \sqrt{n_1  m_3} +  2 \sqrt{n_1 n_2 m_2  n_3} + 2 \sqrt{ n_1 n_2 m_2 m_3 }+
m_2 \sqrt{n_1  n_3} + m_2 \sqrt{n_1  m_3} ) $.
We further see that for non-zero brane-antibrane charges this determinant is non-zero
except for the set of brane-antibrane charges defined by
$ B:=\lbrace  (n1,n2,n3,m1,m2,m3) \mid  n2 \sqrt{m1 n3}+
n_2 \sqrt{m_1 m_3}+ 2 \sqrt{n_2 m_1  m_2 n_3}+ \sqrt{n_2 m_1  m_2 m_3}+
m_2 \sqrt{m_1 n_3}+  m2 \sqrt{m_1 m_3}+ n_2 \sqrt{n_1  n_3} +
n_2 \sqrt{n_1  m_3} +  2 \sqrt{n_1 n_2 m_2  n_3}
+ 2 \sqrt{ n_1 n_2 m_2 m_3 }+ m_2 \sqrt{n_1  n_3} +
m_2 \sqrt{n_1  m_3}=0 \rbrace $ and thus this state-space geometry
is well-defined only on an quotient intrinsic Riemannian manifold
$ N:= M_6 \setminus B $.
The scalar curvature of this state-space manifold may now be computed using
the intrinsic Riemannian metric based on the Hessian matrix of the entropy
corrected by non-extremal contributions.
The exact expression for the scalar curvature is some what involved but
it is seen that the nature of the state-space manifold has similar behavior
to the earlier non-extremal case of charge-anticharge KK-system.
In this case, we may easily observe that the constant entropy curve
is again some non-standard curve which in turn is just given by:
$ (\sqrt{n_1} + \sqrt{m_1}) (\sqrt{ n_2 }+ \sqrt{m_2}) (\sqrt{n_3}+ \sqrt{m_3})
= c $, where $ c $ is a real constant.

We may trace certain behavior of the components of the covariant Riemann
tensors and note for example that the component $ R_{n_1 n_2 m_3 m_4} $
diverges at the roots of the some polynomials of two variables, and
in particular these polynomials as the functions of charge-anticharge of
the branes are $ f_1(n_2,m2)= n_2^4 m_2^3+ 2 (n_2 m_2)^{7/2} + n_2^3 m_2^4 $
and $f_2(n3,m3)= m_3^{9/2} n_3^4 + n_3^4 m_3^{9/2} $.
Whereas some other components of the covariant Riemann tensors may
diverge differently, as for example the $ R_{n3,m3,n3,m3} $ with equal
number of the brane and anti-brane components diverges at the roots of
a single higher degree polynomial: $ f(n_1,m_1,n_2,m_2,n_3,m_3):=
n_2^4 m_2^3 n_3^{9/ 2} m_3^4 + n_2^4 m_2^3 n_3^4 m_3^{9/2}+
2 n_2^{7/2} m_2^{7/2} n_3^{9/2} m_3^4+ 2 n_2^{7/2} m_2^{7/2}n_3^4 m_3^{9/2}+
n_2^3 m_2^4 n_3^{9/2} m_3^4+ n_2^3 m_2^4 n_3^4 m_3^{9/2} $.
We also see that the components of the covariant Riemann tensors may
become zero for certain values of charges-anticharges.
Furthermore, the Ricci scalar curvature diverges at the set
of the roots of the determinant of the metric tensor $ B $,
and becomes null on some single higher degree polynomial,
and exactly at these points in the state-space of the underlying
extremal or near-extremal, general black hole system becomes a
non-interacting statistical system.

Finally, we would like to see the nature of the non-extremal black
hole state-space that what happens to it with the addition of the
KK-monopoles as a non-trivially fibered circle. In the past,
several eminent authors have calculated the entropy for extremal,
near-extremal and general holes which follows the same pattern as
before, for example the case of four charges is considered in
\cite{9603195v1, 9603061v2}: $ S(n_1,m_1,n_2,m_2,n_3,m_3):= 2 \pi
(\sqrt{n_1} + \sqrt{m_1}) (\sqrt{ n_2 }+ \sqrt{m_2})(\sqrt{n_3}+
\sqrt{m_3}) (\sqrt{n_4}+ \sqrt{m_4})$. This leading order entropy
which includes all the special extremal and near-extremal cases
and has been written as a function of charges $ \lbrace n_i
\rbrace $-anticharges $ \lbrace m_i \rbrace $. In this case too
one obtains the same pattern of the underlying state-space
geometry and constant entropy curve as that of three charge
non-extremal black holes but of course, the exact expression for
the geometric invariants such as the determinant of the
state-space metric tensor or that of the scalar curvature are much
more involved. The conclusion to be drawn however remains the same
and as usual the underlying state-space geometry remains
well-defined as certain intrinsic quotient Riemannian manifold $
N:= M_8 \setminus \tilde{B}$, where the set $ \tilde{B} $ is the
set of the roots of the determinant of the metric tensor of
underlying eight charge anti-charges state-space manifold of the
non-extremal black holes with the inclusion of the KK-monopole
contributions.

\section{Multi-centered Black Branes: $D_6D_4D_2D_0$ system}

In this section as a first exercise of the state-space manifold
containing the both single centered black branes and multi-centered black branes,
we study the state-space geometry defined in terms of the extensive
four charges of $D_6D_4D_2D_0$ black brane configuration.
Here we shall explicitly present the analysis of the state-space
geometry arising from the entropy of stationary single-centered
as well as that the multi-centered black hole molecule configurations.
Such multi-centered black hole configurations may be examined by
the so called pin-sized D-brane systems \cite{0705.2564v1, 0702146v2}
and thus we can realize the underlying state-space geometry arising
from the counting entropy of the number of microstates of a zoo
of entropically dominant multi-centered black hole configurations
along with usual single centered black holes.

Although it is well known that the most entropic minimal energy
supergravity solution should be a single centered black hole
\cite{9508072v3}. Thus we are also interested in analyzing the
state-space geometry arising from the leading D-brane entropy with
the leading single centered black brane configurations. But here
our main focus is to study the state-space of the more general
supersymmetric solutions which are visualized as stationary,
multi-centered, molecular black hole bound states determined by
certain electric-magnetic charge centers $ \Gamma_i:=
(p_i^\Lambda, q_{\Lambda,i})$ with respect to certain gauge group
with an index $ \Lambda $ and total charge $ \Gamma= \sum_i
\Gamma_i $, see \cite{0005049v2, 0009234v2} for detailed
constructions of such brane configurations. The main difference
between the single centered black hole configurations and
multi-centered black hole configurations is that the first in the
case of supersymmetric solutions is completely determined by the
topological data of the string compactification while the second
are typically genuine bound states where one can not move the
centers away from each other without supplying some energy to the
system, which is same as some constrained supergravity solutions
\cite{0702146v2, 0304094v1}. For example in the case of two
centered black hole solution, one have equilibrium separation
condition. Moreover, the existence of the multi-centered black
hole bound states depends on the choice of vacuum and in the weak
string coupling limit, the zoo of multi-centered black hole
configurations collapses to a single D-brane.

As the authors of \cite{0705.2564v1, 0702146v2} have shown that in the
suitable parameter regimes, the multi-centered entropy dominates the
single centered entropy in the uniform large charge limit.
We shall thus investigate the state-space geometric implication for
this well studied example of the $D_6D_4D_2D_0$ system.
Consider, a charge $ \Gamma $ obtained by wrapping $D_4,D_2$ and $D_0$
branes around various cycles of a compact space $ X $ which are scaled
up as $\Gamma \rightarrow \Lambda \Gamma $, then there exists a two
centered brane solution with horizon entropy scaling as $ \Lambda^3 $
while that of the single centered entropy scales as $ \Lambda^2 $.
More properly, consider type IIA string theory compactified on a product
of three two-tori $ X:= T_1^2 \times T_2^2 \times T_3^2 $.
Then entropy of a charge $ \Gamma $ corresponding to $ p0 $ $ D_6$
branes on $ X $, $ p $ D branes on $ (T_1^2 \times T_2^2)+(T_2^2 \times T_3^2)
+ (T_3^2 \times T_1^2) $, q $ D_2 $ branes on $ (T_1^2 + T_2^2 + T_3^2) $
and $ q0$ $D_{0} $ branes is given by \cite{0705.2564v1, 0702146v2}:
$ S(\Gamma):= \pi \sqrt{-4 p^3 q0+ 3 p^2 q^2+ 6 p0 pqq0- 4 p0q^3-(p0q0)^2} $.

The metric tensor of the state-space geometry in the entropy representation
may now be obtained as before from the negative Hessian matrix of the entropy
with respect to all the extensive thermodynamic variables which in
this case are just the $D_6, D_4, D_2$ and $D_0$-brane charges.
Explicitly the components of the metric tensor are given as,\\

$g_{p0p0}= -4 \pi \frac{-3p^2q^2q0^2+ 3pq^4q0- q^6+ p^3q0^3}
{(-4 p^3 q0+ 3 p^2 q^2+ 6 p0 pqq0- 4 p0q^3-(p0q0)^2)^{3/2}}$,

$g_{p0p}= 6 \pi \frac{-p^3q0^2q+ 2 p^2q0q^3+ p_2q0^3p0- 2pq^2p0q0^2+ p0q^4q0}
{(-4 p^3 q0+ 3 p^2 q^2+ 6 p0 pqq0- 4 p0q^3-(p0q0)^2)^{3/2}} $,

$g_{p0q}= -12 \pi \frac{2p^3q^2q0+ p^2qq0^2p0- 2pq^3q0p0- q^4p^2+ q^5p0- p^4q0^2}
{(-4 p^3 q0+ 3 p^2 q^2+ 6 p0 p q q0- 4 p0q^3-(p0q0)^2)^{3/2}} $,

$g_{p0q0}= -\pi \frac{6p^4qq0+ 3p^2q^2q0p0- 9pqq0^2 p0^2+ 5q^3 p^3-
6q^4 p0 p+ 6q^3 p0^2 q0+ 6p0 q0^2 p^3+ p0^3q0^3}
{(-4 p^3 q0+ 3 p^2 q^2+ 6 p0 p q q0- 4 p0q^{3}-(p0q0)^2)^{3/2}} $,

$g_{pp}= -12 \pi \frac{p^4q0^2- p^3q^2q0-3p^2qq0p0+ 4pq^3q0p0- p0^2q0^2q^2+ p0^2q0^3p- q^5p0}
{(-4 p^3 q0+ 3 p^2 q^2+ 6 p0 p q q0- 4 p0q^3-(p0q0)^2)^{3/2}} $,

$g_{pq}= 3 \pi \frac{2p^4qq0- 2p0q0^2p^3+ 3p^2q^2q0p0- 3q^3p^3+ 2q^4p0p- pqq0^2p0^2-
2q^3p0^2q0+ (p0q0)^3 }{(-4 p^3 q0+ 3 p^2 q^2+ 6 p0 p q q0- 4 p0q^3-(p0q0)^2)^{3/2}} $,

$g_{pq0}= - 12 \pi \frac{p^5q0- 2p^3q0p0q- p^4q^2+ 2p^2q^3p0+ pq^2p0^2q0- p0^2q^4}
             {(-4 p^3 q0+ 3 p^2 q^2+ 6 p0 p q q0- 4 p0q^3-(p0q0)^2)^{3/2}} $,

$g_{qq}= -12 \pi \frac{4p^3q0p0q- p^2q^3p0- p^2q0^2p0^2- 3pq^2p0^2q0+ p0^2q^4- p^5q0+ p0^3qq0^2}
            {(-4 p^3 q0+ 3 p^2 q^2+ 6 p0 p q q0- 4 p0q^3-(p0q0)^2)^{3/2}} $,

$g_{qq0}= 6 \pi \frac{-p^5q+ 2p^3q^2q0- 2p^2qp0^2+ p0p^4q0- p0^2q^3p+ p0^3q^2q0}
             {(-4 p^3 q0+ 3 p^2 q^2+ 6 p0 p q q0- 4 p0q^3-(p0q0)^2)^{3/2}} $,

$g_{q0q0}= -4 \pi \frac{-p^6+ 3p^4p0q- 3p0^2p^2q^2+ p0^3q^3}
             {(-4 p^3 q0+ 3 p^2 q^2+ 6 p0 p q q0- 4 p0q^{3}-(p0q0)^2)^{3/2}} $.\\

We observe that the determinant of the state-space metric tensor
is $ \Vert g \Vert= 9 \pi^4 $, which is a positive definite
constant. Thus this non-degenerate metric defines a well-defined
state-space which is parameterized in terms of the D-brane charges
$ \lbrace \Gamma:= (p0,p,q,q0) \rbrace $. The scalar curvature
corresponding to the state-space geometry of the equilibrium
$D$-brane microstates may now easily be determined to be
$R(\Gamma)= \frac{8}{3 \pi} (-4 p^3 q0+ 3 p^2 q^2+ 6 p0 pqq0- 4
p0q^3-(p0q0)^2)^{-1/2} $. We observe that the scalar curvature is
a non-zero, positive and everywhere regular function of the
$\Gamma$. This fact seems to be universal and is related to the
typical form for the Ruppeiner geometry as the negative Hessian
matrix of the black brane entropy. As with the standard
interpretation of scalar curvature of the state-space geometry is
to describe the interactions of the underlying statistical system
which thus is non-zero for the $D_6D_4D_2D_0$ black holes. Note
that the absence of divergences in the scalar curvature indicates
this system to be thermodynamically everywhere stable and thus
that there are no phase transitions or any such critical phenomena
in the underlying state-space manifold of this $D$-brane system.
Furthermore, we may easily see that the constant entropy curve is
again some non-standard curve and in turn is given by: $ 4 p^3 q0-
3 p^2 q^2- 6 p0 pqq0+ 4 p0q^3+ (p0q0)^2= c $, which determines the
$D_2D_6D_4D_0$ black brane system in the view-points of
state-space geometry.

Now we shall analyze the associated state-space geometric curvature scalar
for the cases of single and double centered black hole configurations.
Let us take the same cases of charge centers as that of \cite{0705.2564v1, 0702146v2}.
Consider a total charge $ \Gamma:= (p0,p,q,q0)= \Lambda (0,6,0,-12) $ in
a background in which the area of each $ T^2 $ is $ v $, then it is known
that for all $ v $ there exists a single centered black hole solution.
In particular, we see that
$ S( \Gamma= \Lambda (0,6,0,-12))= \pi \sqrt{10368} \Lambda^2 $.
It is easy to find that the scalar curvature remains non zero,
positive and take the value of
$ R( \Gamma= \Lambda (0,6,0,-12))= \frac{\sqrt{2}}{54 \pi \Lambda^2} $.
Further as indicated by Denef and Moore that for $ v > \sqrt{12} \Lambda $,
there exists a two centered bound state with charge centers
$ \Gamma_1 = (1,3 \Lambda, 6 \Lambda^2 ,-6 \Lambda) $
and $ \Gamma_2 = (-1,3 \Lambda, -6 \Lambda^2 ,-6 \Lambda) $.
It is apparent that the two entropy matches for the $ \Gamma_1, \Gamma_2 $
with $ S(\Gamma_1)=  S(\Gamma_2)= \pi \sqrt{108\Lambda^6-36\Lambda^2} \sim \Lambda^3 $
for large $ \Lambda $.
We see further that the state-space scalar curvature again remains non-zero,
positive quantity and takes is the same values for two charge centers and
in particular, it is given by $ R(\Gamma_1)=
R(\Gamma_2)= \frac{4}{9\pi} \frac{1}{\sqrt{3 \Lambda^6-\Lambda^2}} \sim \frac{1}{\Lambda^3} $.
>From this information one may predict that the correlation volume shall
be identical for these two charge centers of a double centered black brane
configurations.

It is instructive to note for large $ \Lambda $ that the state-space curvature
scalar for the case of a single centered black brane configurations goes
as $ R(\Gamma) \sim \frac{1}{\Lambda^2} $,
whereas for the two centered black brane configurations it behaves as
$ R(\Gamma) \sim \frac{1}{\Lambda^3} $.
Therefore, we can analyze the nature of interactions present in the underlying
state-space corresponding to the both single and doubled centered black hole
configurations.
In order to do so, let's consider the limit $\Lambda \rightarrow \infty $
for some fixed  $ v $, then we see that the state-space scalar curvature
of a double centered black hole ceases to zero, more faster than that of
the single centered black hole configuration.
It may further be indicated that the two point correlations becomes very
small in the large charge limit with fixed moduli at infinity and thus
it should indicate a weakly interacting statistical system.
This is indeed correct picture as because the calculation of the entropy
of the $D_6D_4D_2D_0$ black hole is based on the consideration of weakly
bound states of these D-branes over different cycles of the compactifying
compact space $X$.

\section{Fractionation of Branes: Small Black Holes}

Now we analyze the state-space geometry of a given D-brane system from
the perspective of the fractionation of branes and counting the associated
chiral primaries.
For simplicity, let us consider the example of two charge extremal small
black holes in type IIA string theory compactified on $ T^2 \times \mathcal M $,
where $ \mathcal M $ can be either $ K_3 $ or $ T^4 $,
\cite{9707203v1, 0507014v1, 0502157v4, 0505122v2}.
This four dimensional black hole solution is made up out of some $ D_0 $
branes and $ D_4 $ branes wrapping over the $ \mathcal M $.
It is further well-known that this system has near horizon geometry of
$ AdS_2 \times S^2 $, see for details \cite{0005288v1, 9906013v2, 0007195v2, 0004098v2}.

What follows next that in order to make contact of our state-space geometry
with the microscopic perspective, let us consider the chiral primaries of
$ SU(1,1 \mid 2)_Z $ and the associated supersymmetric ground states of
$ \mathcal N= 4 $ supersymmetric quantum mechanics \cite{0412322}.
In this consideration, we may see easily that the are $ 24 p $ bosonic chiral
primaries with total $ D_0 $ brane charge $ N $ in the background with fixed
magnetic $ D_4 $ charge $ p $.
The degeneracy of counting may arise from the combinatorics of the total number
$ N $ of the $ D_0 $ brane charge splitting into $ k $-small clusters with $ n_i $
number of $ D_0 $ branes such that $ \sum_{i=1}^k n_i= N $ on each cluster
corresponding to wrapped $ D_2 $ branes residing on any of the $ 24 p $
bosonic chiral primary states.
Here the counting is done with the degeneracy $ d_N $ of the states having
level $ N $ in a $ (1+1) $ CFT with $ 24 p $ bosons, and thus one renders
at the celebrated microscopic entropy $ S= \ln d_N \simeq 4 \pi \sqrt{Np}=
4 \pi \sqrt{\sum_{i=1}^k n_i p} $. See for more details,
\cite{0409148v2, AtishHarveyPRL, 9511053v1, 0511120v2, 0410076v2}.

Now we shall first explain the state-space geometry for some specific values
of the number of the clusters in which total number $ N $ of the $ D_0 $
brane charge splits.
For k=2 we see that the entropy takes the form, $ S= 4 \pi \sqrt{p(n_1+ n_2)} $.
The components of the associated state-space metric may be easily computed
from the negative Hessian matrix of the fractional brane entropy as,\\

$ g_{p p}= \frac{\pi}{p} \sqrt{\frac{n_1+ n_2}{p}} $,

$ g_{p n_1}= -\frac{\pi}{\sqrt{p(n_1+ n_2)}} $,

$ g_{p n_2}= -\frac{\pi}{\sqrt{p(n_1+ n_2)}} $,

$ g_{n_1 n_1}= \frac{\pi}{(n_1+ n_2)} \sqrt{\frac{p}{n_1+ n_2}} $,

$ g_{n_1 n_2}= \frac{\pi}{(n_1+ n_2)} \sqrt{\frac{p}{n_1+ n_2}} $,

$ g_{n_2 n_2}= \frac{\pi}{(n_1+ n_2)} \sqrt{\frac{p}{n_1+ n_2}} $.\\

Notice that the determinant of the metric tensor is identically zero for
such a microscopic entropy calculation which in turn holds as asymptotic
expansion in the limit of large charges on the branes.
We may further see that the constant entropy curve is given by:
$ p(n_1+ n_2)= c $, for some real constant $c$.
Now we would like to investigate the state-space geometry for three clusters.
In this case, the entropy takes the form, $ S= 4 \pi \sqrt{p(n_1+ n_2+ n_3)} $.
Again one may see that the components of the associated Ruppeiner metric are,\\

$ g_{p p}= \frac{\pi}{p} \sqrt{\frac{n_1+ n_2+n_3}{p}} $,

$ g_{p n_1}= -\frac{\pi}{\sqrt{p(n_1+ n_2+n_3)}} $,

$ g_{p n_2}= -\frac{\pi}{\sqrt{p(n_1+ n_2+n_3)}} $,

$ g_{p n_3}= -\frac{\pi}{\sqrt{p(n_1+ n_2+n_3)}} $,

$ g_{n_1 n_1}= \frac{\pi}{(n_1+ n_2+n_3)} \sqrt{\frac{p}{n_1+ n_2+ n_3}} $,

$ g_{n_1 n_2}= \frac{\pi}{(n_1+ n_2+ n_3)} \sqrt{\frac{p}{n_1+ n_2+ n_3}} $,

$ g_{n_1 n_3}= \frac{\pi}{(n_1+ n_2+ n_3)} \sqrt{\frac{p}{n_1+ n_2+ n_3}} $,

$ g_{n_2 n_2}= \frac{\pi}{(n_1+ n_2+ n_3)} \sqrt{\frac{p}{n_1+ n_2+ n_3}} $,

$ g_{n_2 n_3}= \frac{\pi}{(n_1+ n_2+ n_3)} \sqrt{\frac{p}{n_1+ n_2+ n_3}} $,

$ g_{n_3 n_3}= \frac{\pi}{(n_1+ n_2+ n_3)} \sqrt{\frac{p}{n_1+ n_2+ n_3}} $.\\

We thus again see that in this case too the determinant of the metric tensor
remains zero.
In this case, the constant entropy curve is given by:
$ p(n_1+ n_2+n_3)= c $, for some real constant $c$.
Now let us consider the most general case of the fractional small
black hole microscopic entropy,
$ S= \ln d_N \simeq 4 \pi \sqrt{Np}= 4 \pi \sqrt{\sum_{i=1}^k n_i p} $.
Here again one may find that the components of associated Ruppeiner metric are,\\

$ g_{p p}= \frac{\pi}{p} \sqrt{\frac{\sum_{i=1}^k n_i}{p}} $,

$ g_{p n_i}= -\frac{\pi}{\sqrt{p(\sum_{i=1}^k n_i)}} $, and

$ g_{n_i n_j}= \frac{\pi}{(\sum_{i=1}^k n_i)} \sqrt{\frac{p}{\sum_{i=1}^k n_i}} $,
$ \forall i,j= 1, 2,\ldots, k $.\\

It is again not difficult to see that the determinant of the metric
tensor remains zero for any finite number of clusters corresponding
to wrapped $ D_2 $ branes residing on any of the $ 24 p $ bosonic
chiral primary states.
Hence the state-space geometry of the equilibrium microstates based on the
large charge small black hole entropy obtained from the fractionation
of $ D_0 $ branes is degenerate for any finite numbers of clusters
and thus the corresponding state-space geometry turns out to be ill-defined.
Our conclusion that the state-space geometry of $ N $ fractional branes
with electric charges $ \lbrace n_i \rbrace_{i=1}^{k}$ corresponding to
$ k $ clusters in the background of $ D_4 $ brane with magnetic charge
$ p $ is everywhere ill-defined,
is in very well accordance with the principle of mathematical induction.
As determinant of the metric of state-space geometry is trivially zero
for $ i= 1 $.
We have shown that it remains zero for $ i= 2 $ and $ i= 3 $ and thus
the results hold for $ i= k, \forall k \in Z $.
Further,it is easy to see that the curve:
$ (n_1+ n_2+ \ldots + n_k)p= c $ describes the constant entropy curve
in the state-space of $k$-clustered small black holes.
Therefore, it seems to be certain problems in the application of the OSV
formula to the two charged small black holes \cite{CardosowTalk, 0405146v2}
and thus needs some further investigations.

\section{Mathur's Fuzzball Proposal and Subensemble Theory: Fuzzy Rings}

In this section, we shall study certain aspects of the state-space geometry
for  the case of two charge extremal black holes with an angular momentum
$J$ from the perspective of Mathur's fuzzball proposal and subensemble theory
\cite{0502050v1}.
In this picture one can construct the classical space-time geometry
with a definite horizon when many quanta of the underlying $ D_1 D_5 P $
CFT lie in the same mode.
But in general a generic state will not have its all quanta placed in a few modes,
so the throat of the black hole space-time ends in a very quantum fuzzball
\cite{0706.3884v1, 0109154v1, 0202072v2}.
It is however interesting to note that the actual microstates of the system
in the fuzzball picture do not have horizon,
but it is rather the area of the boundary of the region where microstates start
differing from each other satisfies a Bekenstein type relation with the entropy
inside the boundary.
Moreover, different microstates according to string theory are `cap off' before
reaching the end of the infinite throat and thus give rise to different near horizon
space-time geometries.
In particular, the throat behaves as the inverse of the average radius of fuzzballs.
Thus the Bekenstein entropy may be obtained from the area of the above stretched horizon
with a statistical interpretation as a coarse graining entropy.

\subsection{The Fuzzball Proposal}

It is the associated state-space geometry of the entropy of this rotating two
charge system that we would like to construct whose dimension is equal to the
number of the actual number of the parameters characterizing the brane microstates.
We shall study this state-space of a black ring whose CFT deals with
$ n_1 $ number of $ D_1 $ brane having charge $ Q $ and
$ n_5 $ number of $ D_5 $ brane having charge $ P $.
In the case of two charge extremal holes carrying an angular momentum $J$
which appears quite naturally in the string theory, one finds the stretched
horizon entropy to be $ S(n_1,n_5,J)= c \sqrt{n_1 n_5- J} $. In other words,
as $ Q \sim n_1 $ and $ P \sim n_5 $.
So for some value of $ c $ we may write the ring entropy in terms of the charges
$ Q $ and $ P $ to be $ S(Q,P,J)= C \sqrt{QP- J} $, see for detail
\cite{0706.3884v1, 0109154v1, 0202072v2}.

The state-apace geometry constructed out of the equilibrium state of this
rotating two charged extremal black ring resulting from the entropy may
now be computed as earlier from the negative Hessian matrix of the entropy
with respect to the charges and angular momentum.
Note that the understanding of the state-apace geometry based on the stretched
horizon requires the classical time scale limit of fuzzball.
This is because the size of fuzzball is made by the generic states such that
its surface area in leading order satisfies Bekenstein type relation with the
entropy of the fuzzball whose boundary surface becomes like a horizon only over
classical time scales.
We see that the components of the metric tensor are explicitly given as, \\

$ g_{PP}= \frac{1}{4}CQ^2(PQ-J)^{-3/2} $,

$ g_{PQ}= -\frac{1}{4}C(PQ- 2J)(PQ-J)^{-3/2} $,

$ g_{PJ}= -\frac{1}{4}CQ(PQ-J)^{-3/2} $,

$ g_{QQ}= \frac{1}{4}CP^2(PQ-J)^{-3/2} $,

$ g_{QJ}= -\frac{1}{4}CP(PQ-J)^{-3/2} $,

$ g_{JJ}= \frac{1}{4}C(PQ-J)^{-3/2} $,\\

It is easy to observe as in the previous examples that the Ruppeiner metric is
non-degenerate and regular everywhere. The determinant of the metric tensor is
$ \Vert g \Vert= -\frac{1}{16}C^3 (PQ- J)^{-5/2} $ and thus is non-zero for non-zero charges.
The state-space scalar curvature in the large charge limit in which the asymptotic
expansion of the entropy of the rotating two charge ring is valid, may easily be
computed to be $ R(P,Q,J)= -\frac{5}{2C}(PQ- J)^{-1/2} $.
Notice that the scalar curvature $ R(P,Q,J) $ is a regular function of $ (P,Q,J) $
and thus this system corresponds to everywhere well-defined interacting statistical
system.
It is remarkable that the constant entropy curve in the state-space for any non-zero
rotation takes the form: $ PQ- J= k $, which is just a hyperbolic paraboloid on which
the state-space geometry turns out to be well-defined, interacting statistical system.
Note however that the limit $ J\rightarrow 0 $ makes the state-space geometry to be
ill-defined which in turn is the same case as that of the case of the small black holes.
Now we shall discuss some relations between our state-space geometry and Mathur's
fuzzball proposal of constructing microstates that define a subensemble of the
rotating black holes.

\subsection{Subensemble Theory}

As mentioned in connection with the Mathur's fuzzball proposal that the microstates
of an extremal hole can not have singularity.
Here we find in this picture that even an intrinsic Riemannian manifold of the
equilibrium microstates remains non-singular.
In other words, the regularity of the state-space geometric curvature invariant indicates
the fact that the intrinsic Riemannian geometry constructed out of the microstates of
rotating two charge ring as the maxima of the entropy remains also regular.
The absence of divergences in the scalar curvature consequently implies that the
underlying state-space of rotating two charge ring is thermodynamically stable
and thus there are no phase transitions.
Moreover, the non-zero $ R(P,Q,J) $ indicates that the state-space of the extremal
$D_1D_5J $ system corresponds to an underlying interacting statistical system.

We now indicate the nature of our state-space geometry form the
view-points of the Mathur's subensemble theory. In this
perspective, let us consider the conserved quantities of the ring,
which are here the charges $P$, $Q$ and the angular momentum $J$,
and consider a subset of the states that defines the subensemble.
Let there be some large $M$ number of such subensembles in which
entropy of ring is $\tilde{S}(n_1^{'}, n_5^{'},J)$ for given any
ensemble in which the total entropy of the ring is $S(n_1,n_5,J)$.
Then entropy in each subensemble is given by $ \tilde{S}(n_1^{'},
n_5^{'},J)= \frac{1}{M} S(n_1,n_5,J)$, see for details
\cite{0706.3884v1, 0109154v1, 0202072v2}. Therefore, under the
considerations of the subensemble theory, we can define as before
the state-space geometry to be the negative Hessian matrix of the
ring entropy in any given subensemble with respect to the
conserved charges or number of branes with a rotation. Thus as a
result, we see that the state-space scalar curvature in each
subset gets reduced by some large factor, which in turn is
precisely the number of subensembles. In particular, as $ M
\rightarrow \infty $, we see that $ R \rightarrow 0 $. Therefore
each subensemble with some given number of microstates corresponds
to a non-interacting statistical system in the infinite
subensemble limit. Of course, our this conclusion remains true
only in large charge and large angular momentum limit in which the
computation of ring entropy is valid. But it has been shown in
\cite{0412133v2} that the corrections like space-time higher
derivative terms do not spoils the leading order entropy and thus
the associated state-space geometry.

In the case of the subensemble of a two charge non-rotating extremal hole,
we can easily find that the underlying state-space geometry remains ill-defined,
as the norm of the Hessian matrix of the entropy with respect to the charges
is zero. Thus the state-space geometry in each subensemble of $ D_1 D_5 $
system without angular momentum is ill-defined at the leading order entropy.
But it is known that the higher derivative corrections make the small black hole
state-space geometry well-defined, as there is non-zero subleading entropy
arising from the higher derivative contributions.
What happens really that actually in the limiting picture of angular momentum,
these two rotating and non-rotating state-space geometries will marry each others
in a given ergo-branch.
In conclusion, the state-space geometry of rotating two charged extremal ring in
a given ensemble is curved and corresponds to an interacting statistical system,
whereas it ceases to be non-interacting in each subensemble with given number of
the brane microstates.
However, this fact looses it's meaning for the case of a non-rotating two charge
extremal holes, and the underlying state-space geometry becomes ill-defined in
either an ensemble or in a subensemble of given any ensemble.

\section{Bubbling Black Brane Solutions: Black Brane Foams}

In this section, we analyze the state-space geometry of the equilibrium microstates
of certain foamed black branes having three charges which appears naturally in the
supergravity bubbling solutions, see for details \cite{0604110, 0408120v2, 0409174v2, 0505166v2}.

\subsection{A Toy Model: Single GH-center}

As a first toy model example of the state-space geometry arising from the
entropy of a foam, we consider M-theory compactified on $T^6$ in large N limit,
then one has flux parameters which may be written in terms of brane charges.
Thus there is an associated entropy which is independent of the number and
charges on the Gibbons-Hawking base points.
The origin of this entropy lies solely in the possible number of the choices
of positive quantized fluxes on topologically non-trivial cycles.
These cycles in turn satisfy some constraints, which in particular is that
the supergravity and worldvolume descriptions have the same relations between
the brane parameters which thus determines the entropy of the foam.
In particular, when one considers the flux parameters $ \lbrace k_i^1, k_i^2, k_i^3 \rbrace $
to be positive half integers, the leading order topological entropy coming from
the contributions of $ \lbrace k_i^1 \rbrace $ may be written as \cite{0604110}:
$S(Q_1, Q_2, Q_3):= \frac{\pi}{3}\sqrt{6}(\frac{Q_2 Q_3}{Q_1})^{1/4}$.
In this case, as stated earlier that the state-space metric of the microstates
characterizing the foam is given by the negative Hessian matrix of the foam entropy
with respect to the extensive brane charges and thus is found to be, \\

$ g_{Q_1Q_1}= -\frac{5 \pi \sqrt{6}}{48 Q_1^2} (\frac{Q_2 Q_3}{Q_1})^{1/4} $,

$ g_{Q_1Q_2}=  \frac{\pi \sqrt{6} Q_3}{48 Q_1^2}(\frac{Q_1}{Q_2 Q_3})^{3/4} $,

$ g_{Q_1Q_3}=  \frac{\pi \sqrt{6} Q_2}{48 Q_1^2}(\frac{Q_1}{Q_2 Q_3})^{3/4} $,

$ g_{Q_2Q_2}=  \frac{\pi \sqrt{6}}{16}(\frac{Q_3}{Q_1})^{2} (\frac{Q_1}{Q_2 Q_3})^{7/4} $,

$ g_{Q_2Q_3}= - \frac{\pi \sqrt{6}}{48 Q_1}(\frac{Q_1}{Q_2 Q_3})^{3/4} $,

$ g_{Q_3Q_3}=  \frac{\pi \sqrt{6}}{16}(\frac{Q_2}{Q_1})^{2} (\frac{Q_1}{Q_2 Q_3})^{7/4} $.\\

It may be easily seen that the determinant of the metric tensor is
$ \Vert g \Vert= -\frac{\pi^3 \sqrt{6}}{384 Q_1^4}(\frac{Q_1}{Q_2 Q_3})^{-5/4} $.
Now, one can obtain the associated Christoffel symbols, covariant Riemann tensors,
Ricci tensors in the usual ways, for example the components of the covariant Ricci
tensor are given by, \\

$ R_{11}= \frac{1}{24 Q_1^2} $,

$ R_{12}= -\frac{1}{48 Q_1 Q_2} $,

$ R_{13}=-\frac{1}{48 Q_1 Q_3} $,

$ R_{22}=0 $,

$ R_{23}= \frac{1}{48 Q_2 Q_3} $, and

$ R_{33}= 0 $.\\

Finally, the state-space Ricci scalar can be written to be
$ R= -\frac{1}{2 \pi \sqrt{6}}(\frac{Q_1}{Q_2 Q_3})^{1/4} $.
We see that the Ricci scalar is non-zero and everywhere regular function of the
$ \lbrace Q_1, Q_2, Q_3 \rbrace $ and thus the underlying statistical system
of an unidirectional foamed three charge black brane is everywhere well-defined
and interacting for all $\lbrace Q_1,Q_2,Q_3 \rbrace$.
Further, it is easy to see that the constant entropy curve in the state-space
corresponding to the $k$-direction of the flux parameter is given by the relation:
$ Q_i Q_j= c Q_k $, where $ c $ is a real constant.
Notice further that the same curve describes the constant scalar curvature curve
in the state-space, with some different real constant.

\subsection{Black Brane Foams}

Now we consider the factors coming from the partitioning of all flux parameters
$ \lbrace k_i^1, k_i^2, k_i^3 \rbrace $,
then the leading order topological entropy of the three charged black foam
characterized by the charges $ Q_1 $, $Q_2 $ and $ Q_3 $ is given as \cite{0604110}:
$ S(Q_1, Q_2, Q_3):= \frac{2\pi}{\sqrt{6}}\lbrace (\frac{Q_2 Q_3}{Q_1})^{1/4}
+ (\frac{Q_1 Q_2}{Q_3})^{1/4}+ (\frac{Q_1 Q_3}{Q_2})^{1/4} \rbrace $.
A straightforward computation yields the components of state-space metric tensor
of the foam microstates which in this case are characterized by the three conserved
charges carried by the black brane foam, and in particular are given by,\\

$ g_{Q_1Q_1}= -\pi \lbrace \frac{5 \sqrt{6}}{48 Q_1^2}(\frac{Q_2 Q_3}{Q_1})^{1/4}
- \frac{ \sqrt{6}}{16 Q_1}(\frac{Q_2}{Q_3 Q_1})^{1/4}
- \frac{ \sqrt{6}}{16 Q_1}(\frac{Q_3}{Q_2 Q_1})^{1/4}\rbrace $,

$ g_{Q_1Q_2}= -\pi \lbrace - \frac{ \sqrt{6}Q_3}{48 Q_1^2}(\frac{Q_1}{Q_2 Q_3})^{3/4}
+ \frac{ \sqrt{6}}{48 Q_3}(\frac{Q_3}{Q_1 Q_2})^{3/4}
- \frac{ \sqrt{6}Q_3}{48 Q_2^2}(\frac{Q_2}{Q_1 Q_3})^{3/4} \rbrace $,

$ g_{Q_1Q_3}= -\pi \lbrace - \frac{ \sqrt{6}Q_2}{48 Q_1^2}(\frac{Q_1}{Q_2 Q_3})^{3/4}
- \frac{ \sqrt{6}Q_2}{48 Q_3^2}(\frac{Q_3}{Q_1 Q_2})^{3/4}
+ \frac{ \sqrt{6}}{48 Q_2}(\frac{Q_2}{Q_1 Q_3})^{3/4} \rbrace $,

$ g_{Q_2Q_2}= -\pi \lbrace \frac{5 \sqrt{6}}{48 Q_2^2}(\frac{Q_1 Q_3}{Q_2})^{1/4}
- \frac{ \sqrt{6}}{16 Q_2}(\frac{Q_3}{Q_1 Q_2})^{1/4}
- \frac{ \sqrt{6}}{16 Q_2}(\frac{Q_1}{Q_3 Q_2})^{1/4}\rbrace $,

$ g_{Q_2Q_3}= -\pi \lbrace \frac{ \sqrt{6}}{48 Q_1}(\frac{Q_1}{Q_2 Q_3})^{3/4}
- \frac{ \sqrt{6}Q_1}{48 Q_3^2}(\frac{Q_3}{Q_1 Q_2})^{3/4}
- \frac{ \sqrt{6}Q_1}{48 Q_2^2}(\frac{Q_2}{Q_1 Q_3})^{3/4} \rbrace $,

$ g_{Q_3Q_3}= -\pi \lbrace \frac{5 \sqrt{6}}{48 Q_3^2}(\frac{Q_1 Q_2}{Q_3})^{1/4}
- \frac{ \sqrt{6}}{16 Q_3}(\frac{Q_2}{Q_1 Q_3})^{1/4}
- \frac{ \sqrt{6}}{16 Q_3}(\frac{Q_1}{Q_2 Q_3})^{1/4}\rbrace $.\\
\\
The determinant now takes simple form
$ \Vert g \Vert= -\frac{\pi^3 \sqrt{6}}{384} (Q_1 Q_2 Q_3)^{-13/4} f_1(Q_1,Q_2,Q_3) $,\\
\\
where $ f_1 $ is defined by
$ f_1(Q_1,Q_2,Q_3):= - Q_1^{3/2} Q_2 Q_3^{2} - Q_1 Q_2^{3/2} Q_3^{2}
+ 3 Q_1^{3/2} Q_2^{3/2} Q_3^{3/2}- Q_1^{3/2} Q_2^{2} Q_3
- Q_1 Q_2^{2} Q_3^{3/2} - Q_1^{2} Q_2^{3/2} Q_3
+ Q_1^{2} Q_2^{1/2} Q_3^{2} + Q_1^{2} Q_2^{2} Q_3^{1/2}
- Q_1^{2} Q_2 Q_3^{3/2} + Q_1^{1/2} Q_2^{2} Q_3^{2} $.
We thus observe a non-degenerate metric for the state-space geometry at extremality
as an intrinsic quotient Riemannian manifold.
In this case again the $\Gamma_{ijk}, R_{ijkl}, R_{ij}$ are not difficult to obtain.
The curvature scalar in the state-space geometric framework may easily computed to be,\\

$ R= -\frac{\sqrt{6}}{12 \pi} (Q_1 Q_2 Q_3)^{7/4} f_2(Q_1,Q_2,Q_3) f_1(Q_1,Q_2,Q_3)^{-3} $,
where $ f_2 $ is defined by \\

$ f_2(Q_1,Q_2,Q_3)= -10 Q_1^{4} Q_2^{2} Q_3^{2}- 10 Q_1^{2} Q_2^{2} Q_3^{4}- 10 Q_1^{2} Q_2^{4} Q_3^{2}
+ Q_1^{4} Q_2^{4} + Q_1^{4} Q_3^{4} + Q_2^{4} Q_3^{4}
+ 4 Q_1^{4} Q_2^{5/2} Q_3^{3/2}- 27 Q_1^{3} Q_2^{2} Q_3^{3}+ 4 Q_1^{5/2} Q_2^{5/2} Q_3^{4}
+ 4 Q_1^{3/2} Q_2^{5/2} Q_3^{4}+ 4 Q_1^{5/2} Q_2^{4} Q_3^{3/2}- 4 Q_1^{4} Q_2^{7/2} Q_3^{1/2}
+ 4 Q_1^{3} Q_2^{4} Q_3- 4Q_1^{7/2} Q_2^{4} Q_3^{1/2}- 27 Q_1^{3} Q_2^{3} Q_3^{2}
-4 Q_1^{1/2} Q_2^{7/2} Q_3^{4}- 4Q_1^{2} Q_2^{7/2} Q_3^{5/2}+ 6Q_1^{3} Q_2^{3/2} Q_3^{7/2}
-4 Q_1^{7/2} Q_2^{2} Q_3^{5/2}- 4Q_1^{7/2} Q_2^{5/2} Q_3^{2}+ 22 Q_1^{5/2} Q_2^{5/2} Q_3^{3}
-4 Q_1^{5/2} Q_2^{7/2} Q_3^{2}
-4 Q_1^{5/2} Q_2^{2} Q_3^{7/2}+ 2Q_1 Q_2^{7/2} Q_3^{7/2}+ 2 Q_1^{7/2} Q_2 Q_3^{7/2}
+2 Q_1^{7/2} Q_2^{7/2} Q_3+ 22 Q_1^{5/2} Q_2^{3} Q_3^{5/2}+ 6 Q_1^{7/2} Q_2^{3/2} Q_3^{3}
+ 4 Q_1^{4} Q_2 Q_3^{3}- 4Q_1^{7/2} Q_2^{1/2} Q_3^{4}+ 4 Q_1^{3} Q_2 Q_3^{4}
+ 4 Q_1 Q_2^{4} Q_3^{3}+ 4 Q_1^{3/2} Q_2^{4} Q_3^{5/2}+ 4 Q_1^{4} Q_2^{3/2} Q_3^{5/2}
+ 4 Q_1 Q_2^{3} Q_3^{4}+ 4 Q_1^{4} Q_2^{3} Q_3- 4 Q_1^{1/2} Q_2^{4} Q_3^{7/2}
-27 Q_1^{2} Q_2^{3} Q_3^{3}+ 6 Q_1^{7/2} Q_2^{3} Q_3^{3/2}+ 6Q_1^{3} Q_2^{7/2} Q_3^{3/2}
+22 Q_1^{3} Q_2^{5/2} Q_3^{5/2}- 4Q_1^{2} Q_2^{5/2} Q_3^{7/2}
+ 6 Q_1^{3/2} Q_2^{3} Q_3^{7/2}+ 6 Q_1^{3/2} Q_2^{7/2} Q_3^{3}- 4 Q_1^{4} Q_2^{1/2} Q_3^{7/2} $.\\

The curvature scalar is thus finite and everywhere regular for all well-defined foam
charges $\lbrace Q_1,Q_2,Q_3 \rbrace$, except for the case of $ f_1(Q_1,Q_2,Q_3)= 0 $.
It is however interesting to note that in the large charge limit at which the entropy
computation is valid, we see that whenever $ f_2(Q_1,Q_2,Q_3)\neq 0 $,
the underlying scalar curvature is a non vanishing function of the brane charges,
and thus indicates an underlying interacting statistical system.
We may easily see that the underlying state-space geometry is well-defined
only as an intrinsic quotient Riemannian manifold,
$ M:= M_3 \setminus B $, where the set of charges $ B $ is defined by
$ B:= \lbrace (Q_1,Q_2,Q_3) \vert f_1(Q_1,Q_2,Q_3)= 0 \rbrace $.
Moreover, we observe that in this case for some real constant $ c $,
the constant entropy curve is defined by the equation:
$ (\frac{Q_2 Q_3}{Q_1})^{1/4}+ (\frac{Q_1 Q_2}{Q_3})^{1/4}+ (\frac{Q_1 Q_3}{Q_2})^{1/4}= c $.
We see easily that in this case it is not the same case as that of the previous
subsection but rather here the curve of constant curvature scalar is given by
$ f_1(Q_1,Q_2,Q_3)= K f_2(Q_1,Q_2,Q_3) $.
Furthermore, this system with non-zero three flux parameters corresponds to
a non-interacting statistical system on a hypersurface defined by the equation
$ f_2(Q_1,Q_2,Q_3)= 0 $.
Note however that the entropy considered here is the one arising from the
usual two derivative terms in the low energy effective supergravity action
which is consistent with the area law.
Certainly, the entropy expression will be modified by contributions from the
higher derivative terms in the effective action or the corrections arising
from the topological Chern numbers, see for example \cite{0506015v2} in the
case of microscopic description.
This would consequently modify the equilibrium microstates and thus the
corresponding state-space geometry including the curvature scalar.

In this section we have thus explored the state-space geometry of a large family
of smooth three charged BPS space-time geometries.
We find that in the case of large number of two cycles, a large class of these
black brane foams having the correct charges and angular momenta defines a
non-degenerate, intrinsic Riemannian geometry which may realize the correlations
among the microstates of a three charged maximally spinning BPS black hole in
five dimensions just by calculating the equilibrium state-space metric tensor
and the associated scalar curvature.
It is well known that this solution can be dualized to a frame of the $D_1 D_5 P$
charges which is asymptotic to $ AdS_3 \times S^3 \times T^4 $ \cite{0408186v2}.
Hence we can have some clue of the two point correlation functions of the dual
$ D_1 D_5 P$-CFT states.
This may allow us to find whether the typical correlation functions of black brane
CFT microstates are dual to the thermodynamic correlation functions arising from
the Gaussian fluctuations of the topological entropy of bubbling brane solutions.

\section{Discussion and Conclusion}

In this paper, we have discussed state-space geometry of various interesting
cases of string theory extremal and non-extremal black brane solutions
with or without higher derivative $\alpha^{\prime}$-corrections,
multi-centered black brane configurations, fuzzy rings and
bubbling black brane foam solutions in $ M $-theory.
We find in all such cases that the underlying state-space geometry is
well-defined and everywhere regular, intrinsic (quotient) Riemannian manifold.
For instance, the simplest example analyzed is the case of $R^4$-corrected
non-supersymmetric extremal small black holes in four dimensions for which
the state-space is everywhere well-defined and corresponds to a non-interacting
statistical system.
The fact that the various state-space geometry turned out to be everywhere regular
is in perfect agreement with another fact that such BPS black branes are
stable objects even under the consideration of higher derivative corrections,
or brane fractionation, or Mathur's theory of subensemble,
or $ M $-theory bubbling black brane configurations.

We have in turn investigated the state-space geometry arising from the entropy
of the extremal black branes as a bound states of large number of microstates
having certain number of electric-magnetic charges and angular momenta
as well as that of the mass associated with the non-extremal black brane
configurations.
In the first case of an stringy extremal black hole whose microstates are just
the winding and momentum modes of a heterotic string carrying $ n_1 $
number of winding and $ n_p $ number of momentum leads to the large
charge microscopic entropy $ S_{micro}= 2\sqrt{2n_1 n_p} $.
Macroscopically, the entropy of this two charged black hole may be computed
by considering $D_4$ and $D_0$-branes in a specific compactification such as
$T^6$ or $K_3 \times T^2$ down to the usual Minkowskian space-time, then one
finds that leading order entropy vanishes.

But the higher order stringy corrections like the $ R^2$-corrections modify the
usual Einstein-Hilbert action and then with such corrections one finds that
the horizon area is being stretched which in turn leads to a non-zero black
brane entropy.
The microscopic calculation of the entropy for $n_0$ number of $D_0$-branes
and $n_4$ number of $D_4$-branes deals with the counting of the weakly
interacting D-brane ensemble and in particular, it turns out that
$ S_{micro}= 2 \pi \sqrt{n_0 n_4}= S_{macro} $.
We have here shown that the state-space geometry associated with this system is ill-defined
which may easily be seen from the determinant of the negative Hessian matrix of the entropy
with respect to the charges carried by the branes.
Further, this result holds for any constant entropy curve whose state-space coordinates
lies on a rectangular hyperbola.

However, on the inclusion of the $ R^4 $-higher derivative corrections in the space-time
for the case of the non-supersymmetric two charge small black holes,
we find that this state-space geometry becomes a well-defined intrinsic Riemannian geometry.
A more closure look of such corrected state-space geometry of the equilibrium
microstates for the two charged small black holes reveals to the conclusion that
the state-space of non-supersymmetric two charge small black holes with quartic
corrections is a non-interacting statistical system
and thus it indicates that there are no thermodynamic instabilities
and phase transitions anywhere in this state-space whose constant
entropy curve takes the form of a hyperbola.
Moreover, the similar results hold for three and four charged tree
level extremal black holes whose state-space geometry takes the same
form of the associated metric tensors and in turn yields everywhere regular scalar curvatures
corresponding to the charged extremal black hole configurations.
In these cases, it turns out that the determinant of the metric tensors and
the scalar curvatures both scale as the inverse of the square root of the leading order
brane entropies and the constant entropy curve in the associated state-space takes
the form of some higher dimensional hyperbola: $\prod_i n_i = c^2 $.

The state-space geometry for the non-extremal black holes turns out to be
more interesting from the view-points of the underlying interactions that
are present among the bound states of the brane-antibrane pairs.
As a non-extremal black hole configuration is obtained by adding the
corresponding anti-branes to the extremal black brane configurations.
In particular, the microscopic description of a simplest non-extremal $D_1, D_5$
system requires a string with some extra energy having large amount of winding
and $D_5$ brane charges which in turns creates the equal amount of momenta
running in the opposite direction of the $ S^1$.
In this case the state-space metric tensor computed as the negative Hessian matrix
of the entropy implies a non-zero determinant of the metric tensor
for any set of given non-zero brane-antibrane charges.
In particular, we find that the state-space geometry of the
four charged non-extremal $ D_1 D_5 P $ black hole system
is non-degenerate.
The curvature scalar with such non-extremal contributions turns out to be non-zero,
except for a set of the roots of a two variable polynomial $ f(n_p, \overline{n_p}) $
as a function of the two momenta running in the opposite directions of the $S^1$.
This polynomial in the state-space may easily be seen to be the function of the charges
and anti-charges running on the $ S^1$ which in turn yields an everywhere regular intrinsic
Riemannian manifold and thus corresponds to an interacting statistical system,
except at the roots of the $ f(n_p, \overline{n_p}) $ on which the underlying
statistical system becomes a non-interacting statistical system.
In this case we find that the constant entropy curve is no more a standard
hyperbola as that of the extremal black branes but it is given to be
$ \frac{c^2}{n_1 n_5}= (\sqrt{n_p}+ \sqrt{ \overline{n_p}})^{2} $.

One can further extrapolate the above leading order entropy to a non-large charge domain,
where the microstates are no longer close to the supersymmetric states
and thus deals in one stroke with all the special extremal and near-extremal
configurations.
For example, it is not difficult to explore the state-space geometry
arising from the leading order entropy expression of the equilibrium
microstates of the 6 charge anti-charge non-extremal black hole in $D=4$.
We see that state-space geometry defined as the negative Hessian matrix
of the entropy with respect to the conserved charges-anticharges
carried by the black hole is non-degenerate for all non-zero brane-antibrane
charges except for a set $B$ of brane-antibrane charges defined by the
roots of the determinant of the state-space metric tensor.
Thus this state-space geometry remains well-defined only as an intrinsic
quotient Riemannian manifold $ N:= M_6 \setminus B $.
The scalar curvature in this case may further easily be computed by using
the intrinsic Riemannian metric as the negative Hessian matrix of the corrected entropy
with the non-extremal contributions
and in turn we have shown that the state-space geometry has the similar nature
to that of the non-extremal charge-anticharge KK system.
It is worth to mention that in this case the constant entropy curve is a non-standard curve:
$ (\sqrt{n_1} + \sqrt{m_1}) (\sqrt{ n_2 }+ \sqrt{m_2}) (\sqrt{n_3}+ \sqrt{m_3})
= c $, where $ c $ is a real constant.

It is worth to note that some components of the covariant Riemann tensors may
diverge at the roots of the some polynomials of lower degrees as the functions
of the charges-anticharges, while those components of the covariant Riemann
tensors which comes with the equal brane-antibrane components diverge differently
as the roots of a single higher degree polynomial.
We also observe that certain components of the covariant Riemann tensors may
become zero for certain values of charges-anticharges.
Furthermore, it is not difficult to see that the Ricci scalar curvature diverges
on the set $ B $ of the roots of the determinant of the metric tensor,
as well as it becomes null on some single higher degree polynomial,
and exactly at these points in the state-space of the underlying
extremal or near-extremal, or that of a general black hole system becomes
a non-interacting statistical system.

It is important to note further that the nature of the non-extremal black hole
state-space with the addition of the KK-monopoles as a non-trivially fibered circle
follows the same pattern as before as a function of charges-anticharges.
In this case too, we see that there exists the same pattern of the
state-space geometry and that the constant entropy curve retains the similar
expression to that of the three charge non-extremal black holes.
The conclusion to be drawn thus remains the same that the underlying
state-space geometry is well-defined as an intrinsic quotient Riemannian
manifold $ N:= M_8 \setminus \tilde{B}$, where the set $ \tilde{B} $ is
defined as the set of the roots of the determinant of the metric tensor
of the underlying state-space geometry, which in turn is characterized by
the charges and anti-charges carried by the non-extremal black hole.

The state-space geometry containing the four charged single centered black brane configurations
and that of the multi-centered black brane molecule configurations may be examined
under the considerations of the pin-sized $D$-brane systems.
In particular, it turns out that the state-space geometry arising from the counting entropy
of the microstates of the entropically dominant multi-centered $D_6D_4D_2D_0$ black holes
is weakly interacting statistical system in contrast to the case of the single centered black holes.
Note that this observation is in turn consistent with fact that the existence of the
multi-centered black hole bound states depends on the choice of vacuum, and
such multi-centered black hole configurations in the weak string coupling limit
collapse to a single D-brane configuration.
However, we see that in the suitable parameter regimes, the uniform large charged
multi-centered configurations dominate the single centered configurations.
We have shown that the state-space geometric implication for the case of
$D_6D_4D_2D_0$ system with a charge vector $ \Gamma $ obtained by wrapping
$D_4,D_2$ and $D_0$ branes around various cycles of a three two-tori
$ X:= T_1^2 \times T_2^2 \times T_3^2 $ on which the type IIA string theory
is being compactified indicates that there exists two regions in the state-space
which respectively deal with a single centered and multi-centered solutions
for a scaling $\Gamma \rightarrow \Lambda \Gamma $.
In particular, the interactions present in the state-space for the case of the two
centered solutions scale as $ \Lambda^{-3} $ while they scale as $ \Lambda^{-2} $
for the single centered D-brane configurations.

It turns out that the state-space geometry in the entropy
representation obtained from the negative Hessian matrix of the
entropy with respect to the $D$-brane charges has a positive
definite constant determinant and thus this state-space defines a
well-defined non-degenerate intrinsic Riemannian geometry, which
in turn is parameterized by the D-brane charges $ \lbrace \Gamma:=
(p0,p,q,q0) \rbrace $. We have observed that the corresponding
scalar curvature of this state-space manifold is non-zero,
positive and everywhere regular function of the D-brane charges.
Moreover, this fact seems to be universal which is related to the
typical form for the state-space geometry arising from the
negative Hessian matrix of the black brane entropy. As with the
standard interpretation of the state-spaces geometry, the
non-vanishing of the scalar curvature describes that the
underlying statistical system is in general interacting for the
$D_6D_4D_2D_0$ configurations. Note that the absence of the
divergences in the scalar curvature indicates that this system is
thermodynamically everywhere stable and thus there are no any
phase transitions or any such critical phenomena in the
state-space manifold of this $D$-brane system whose constant
entropy curve is given by: $ 4 p^3 q0- 3 p^2 q^2- 6 p0 pqq0+ 4
p0q^3+ (p0q0)^2= c $.

It is worth to mention that the case of a single centered black hole configuration
and that of the double centered black hole configurations,
which are respectively described by a total charge center $ \Gamma:= \Lambda (0,6,0,-12) $
and the two charge centers $ \Gamma_1 = (1,3 \Lambda, 6 \Lambda^2 ,-6 \Lambda)$ and
$ \Gamma_2 = (-1,3 \Lambda, -6 \Lambda^2 ,-6 \Lambda) $,
we find that the state-space scalar curvatures remains non-zero, positive quantities
and in particular, take the same numerical values for the case of the two charge centers
$ \Gamma_1$ and $\Gamma_2$.
Therefore, we predict from this information that the correlation volumes remain
identical in the two sub-configurations of a double centered black brane configuration.
It is thus instructive to note that the nature of interactions present in the underlying
state-space manifold corresponding to the single and doubled centered configurations
behaves differently.
In particular, we see that the state-space scalar curvature of the double centered
configurations ceases to zero more faster than that of the single centered configurations
in the limit of $\Lambda \rightarrow \infty $.
It may further be indicated that the two point correlations becomes very
small in the uniform large charge limit with fixed moduli at the infinity,
and thus they both should indicate a weakly interacting statistical system.
This is indeed the correct picture as the calculation of the entropy is based
on the consideration that such a brane configuration is a bound states of the
weakly interacting D-branes.

On other hand, in order to make contact of our state-space
geometry with the brane fractionation, we have considered the
chiral primaries of $ SU(1,1 \mid 2)_Z $ in a supersymmetric
ground states of $ \mathcal N= 4 $ supersymmetric quantum
mechanics of the D-branes associated with the two charge extremal
small black holes having $ AdS_2 \times S^2 $ near horizon
geometry which are obtained in the compactification of type IIA
string theory either on $ K_3 \times T^2 $ or that of on $ T^6 $.
In particular, it is known from the perspective of fractionation
of branes that this consideration deals with the $ D_0, D_4
$-branes wrapping either $ K_3 $ or $ T^4 $ where the $ 24 p $
bosonic chiral primaries with total $ N $ unit of $ D_0 $-brane
charge in the background of the $ D_4 $-brane with $ p $ unit of
the magnetic charges implies that the counting degeneracy arising
from the combinatorics of the total charges on the $ D_0 $-brane
splitting into $ n_i $ number of $ k $-small clusters such that on
each cluster corresponding to wrapped $ D_2 $-branes residing on
either of the $ 24 p $ bosonic chiral primary states satisfying $
\sum_{i=1}^k n_i= N $ corresponding to the states at the fixed $
N^{th} $-level in an underlying $ (1+1) $-CFT renders to the
celebrated microscopic entropies in either cluster. The
state-space geometry thus computed from the negative Hessian
matrix of such a fractional brane entropy for any arbitrary finite
number of the clusters in which the total $ N $ unit of the $ D_0
$-brane charge splits turns out to be a degenerate intrinsic
Riemannian manifold. Further, it is easy to see that this
ill-defined state-space geometry of $ N $ fractional branes with $
\lbrace n_i \rbrace_{i=1}^{k}$ electric charges corresponding to
the $ k $-clusters in the background of $ D_4 $-branes having $ p
$ unit of magnetic charges may be defined by the curve: $ (n_1+
n_2+ \ldots + n_k)p= c $ which in fact describes the constant
entropy curve in the state-space of $k$-clustered small black
holes.

In order to acquire further supports with the recent microscopic studies of black hole physics,
we have investigated that the state-space geometry of two charge extremal black holes with
an angular momentum renders to be very illuminating from the perspective of the fuzzballs
and subensemble theory.
In particular, we have considered many quanta of the underlying $ D_1 D_5 P $-CFT
lying in the same mode which defines the classical black hole space-time geometry
with a definite horizon.
As in general for a generic state all the quanta may not be placed in a few modes,
so the throat of the black hole space-time ends in a very quantum fuzzball.
Further with the fact that the actual microstates of this system do not have horizon,
but it is the boundary region of the microstates where they start differing
from each other and the entropy enclosed inside the surface area of this boundary
as a coarse graining statistical interpretation
satisfies a Bekenstein-Hawking type relation which in turn
leads to the fact that there exists a non-singular intrinsic Riemannian manifold
of the equilibrium microstates in the fuzzball theory.
This is in perfect accordance with the ideas of the stretched horizon
which arises from the fact that the different microstates are caped off
before reaching the end of the infinite throats and thus
have different near horizon space-time geometries.

In particular, we have shown that
the state-space constructed out of the equilibrium microstates
of an extremal black ring whose CFT deals with
$ n_1 $ number of $ D_1 $ brane having charge $ Q $ and
$ n_5 $ number of $ D_5 $ brane having charge $ P $
carrying an angular momentum $J$
is a non-degenerate, everywhere regular intrinsic Riemannian manifold.
This is due to fact that the determinant of the metric tensor is non-zero
and the state-space scalar curvature in the large charge limit in which the asymptotic
expansion of the entropy of the rotating two charge ring is valid turns out to be
a non-zero, regular function of $ (P,Q,J) $ which is in perfect accordance
with the Mathur's fuzzball proposal that the microstates
of an extremal hole can not have singularity.
The absence of the divergences in the scalar curvature consequently implies that the
underlying state-space of rotating two charge ring is thermodynamically stable
and thus there are no phase transitions.
As a consequences, this system corresponds to everywhere well-defined interacting statistical system
whose constant entropy curve in the state-space for any non-zero rotation takes the form
of a hyperbolic paraboloid on which the state-space geometry is well-defined.
It is worth to mention that the underlying state-space geometry turns out be ill-defined
in the vanishing angular momentum limit
which in turn is the same case as that of the small black holes.

We have further indicated the nature of our state-space geometry of such a ring
form the perspective of Mathur's subensemble theory.
In particular, we considered the conserved charges and angular momentum
of the ring and focused our attention in a subset of the microstates that
defines the subensembles such that the total ring entropy as an additive thermodynamic
quantity remains conserved.
Under this consideration, we find as before that the state-space geometry
defined in any given subensemble as the negative Hessian matrix of the ring entropy
in the large charge and large angular momentum limit
with respect to the conserved charges or the number of branes
to be a non-degenerate and everywhere regular, curved, intrinsic Riemannian manifold.
Moreover, we see that the state-space scalar curvature in each subset
gets precisely reduced by the number of the subensembles.
In particular, each subensemble with large  number of given microstates corresponds
to a non-interacting statistical system in the infinite subensemble limit.
It is worth to mention that the higher derivative space-time corrections do not spoils
our state-space geometric conclusions as the associated extremal ring entropy remains
intact under such corrections, see for details \cite{0412133v2}.

In the case of vanishing angular momentum, we have demonstrated that
the state-space geometry under any subensemble of a two charge non-rotating extremal holes
is ill-defined.
This may easily be read off from the norm of the state-space Hessian matrix of
the entropy with respect to the brane charges.
Thus the state-space geometry in each subensemble of $ D_1 D_5 $-CFT
without angular momentum is ill-defined at the leading order entropy.
However, in this case it is known that the further higher derivative contributions
are nontrivial and thus the non-zero subleading small black hole entropy
corresponds to a well-defined state-space geometry.
One interesting aspect of our geometric study is that in a given ergo-branch
these two rotating and non-rotating state-space geometries
will marry each others in the limiting picture of angular momentum
in either an ensemble or any subensemble of the given ensemble.

As a final exercise, we have analyzed the state-space geometry
of a three charged foamed bubbling black brane supergravity configurations
from the perspective of the large N limit of M-theory compactified on the $T^6$.
In particular, the origin of the entropy of the foam
is independent of the number of charges and Gibbons-Hawking base points
but it solely lies on the number of possible choices of positive
quantized fluxes on each topologically non-trivial cycles
defining the extensive brane charges in terms of the
associated flux parameters to be either positive half integers
or positive integers.
We have shown that the state-space geometry arising from the negative
Hessian matrix of the foam entropy associated with the single center
Gibbons-Hawking base points with respect to the extensive brane charges
is a non-degenerate, everywhere regular, curved intrinsic Riemannian manifold.
In this case, it turns out that the underlying statistical system
of such a three charge unidirectional black foamed is everywhere well-defined
and is an interacting statistical system for the extensive brane charges.
Furthermore, it is shown up to the scaling
that the both constant entropy curve and constant state-space curvature curve
are defined by the constraint $ Q_i Q_j= c Q_k $, where the
non-zero flux lies in the $k$-direction of the state-space of
the three charged black foam.

In the most general consideration,
when all the flux factors coming from various partitioning of all flux parameters
contribute to the leading order topological entropy of the three charged black brane foam,
we have shown that the state-space geometry characterized by the entropy as a function
of the three conserved charges of the black foam pertains to be a non-degenerate intrinsic
Riemannian manifold, except for the case of $ f_1(Q_1,Q_2,Q_3)= 0 $.
In this case further, it turns out that the state-space scalar curvature
is everywhere regular and finite function of the foam charges
and thus it indicates an underlying interacting statistical system,
except for the case of $ f_2(Q_1,Q_2,Q_3)= 0 $.
Furthermore, it is worth to mention that the underlying state-space geometry
is defined by: $ (\frac{Q_2 Q_3}{Q_1})^{1/4}+ (\frac{Q_1 Q_2}{Q_3})^{1/4}+
(\frac{Q_1 Q_3}{Q_2})^{1/4}= c $ constant entropy curve and in turn
remains well-defined as an intrinsic quotient Riemannian manifold $ M:= M_3 \setminus B $,
where the set of charges $ B $ is defined as a set of the roots of the
determinant of the state-space metric tensor.
In this case it is easy to see that it is not the same case as that of the
case of a three charge unidirectional black brane foam
but rather the constant scalar curvature curve turns out to be a non-trivial constraint
on the state-space $ f_1(Q_1,Q_2,Q_3)= K f_2(Q_1,Q_2,Q_3) $ for some $K\in R$.
It is worth to emphasize that the state-space of a three charge black foam with
three non-zero flux parameters becomes a non-interacting statistical system on
a hypersurface $ f_2(Q_1,Q_2,Q_3)= 0 $, whereas that of the three charge unidirectional
black foam is always an interacting statistical system.

Note however that the state-space geometry of the  black brane foams thus considered
is the one arising from the entropy obtained from the leading order contributions
of a low energy effective supergravity action
but certainly the entropy expression will be modified by the contributions coming from the
higher derivative corrections arising from topological Chern numbers
which would consequently modify the equilibrium state-space geometry of the black foams.
However, the state-space geometries explored for a large family of smooth three charged BPS
space-time geometries indicates that the nature of the correlation volume and that of the
two point correlation functions among the microstates of a three charged maximally spinning
BPS black hole in five dimensions as well as that of a large class of black brane foams
having the correct charges and angular momenta may easily be analyzed in the case of
large number of two cycles.
It is further instructive to investigate in more details whether such correlations
in the state-space arising from the Gaussian fluctuations of the topological entropy
of the bubbling black brane configurations are the same as that of those obtained from
the corresponding dual $ D_1 D_5 P$-CFT's.

It is worth to note however that the construction of any state-space manifold
whose constant entropy curve is a $k$-rectangular hyperbola in a given $S,T$-duality
basis is ill-defined.
This in turn is the case of the state-space of the extremal small holes with $k$-clusters
of the $ D_0 $-branes.
However, the underlying state-space may be made to be well-defined by adding
one more coordinate to the constant entropy curve in the state-space.
The simplest example of this case that we have demonstrated in detail is
the fuzzy black ring whose constant entropy curve in the  state-space is
just the hyperbolic paraboloid.
Therefore a further understanding of these matters will ultimately depend on
how well the entropy of a given brane configuration can be understood.
>From the perspective of Weinhold geometry, these effects are directly
related to the moduli space geometry and thus one may try from a different
view-point to understand the microscopic origin of thermodynamic geometry
of extremal as well as non-extremal black branes arising from the
consideration of the large number of coincident D-branes or M-branes.

In other words, the study of covariant state-space geometry thus seeds light
on the nature of the two point correlation functions and correlation volume
of the underlying boundary conformal field theory;
whether one would have a stable or unstable system from the view-points of a CFT.
Therefore, it seems possible to understand a microscopic origin of thermodynamic
singularities of the non-extremal black branes from the associated CFT data consisting
of brane-antibrane pairs.
In short, our geometric notions are important to understand the nature of the
interactions present in the dual CFT of various extremal as well as non-extremal
black holes and black branes arising from certain compactifications of either a
string theory as well as that of a M-theory or vice-versa.

Finally, in certain cases of our study of the state-space geometry,
it is however found that the determinant of the state-space metric tensor is negative,
whereas in some cases we find it to be positive definite.
In particular, the foamed three charge black branes and fuzzy black rings
has a negative determinant where as it is everywhere positive for the two
centered $D_6D_4D_2D_0$ black holes.
In the case of specific D-brane solutions, it seems thus that there is an
attractor tree and topological flow tree data but it is not the case with
the charged black brane foams or fuzzy rings and thus their entropy needs
some further corrections, which may possibly for example be the case of the
non-perturbative instanton corrections.
In conclusion, we have shown in all the cases considered in this article that
there are no singularity in the state-space geometry of the black brane systems.
Our geometric study of state-space thus supports the general conjecture
that the interior of such black branes is non-trivial and there are no
singularities in the state-space except those that arrive at the roots of
the determinant of the metric tensor of the associated state-space geometry.
It thus renders to the conclusion that such brane state-space geometries
are in general non-degenerate, completely smooth, (quotient) intrinsic Riemannian
manifolds for all well-defined charges of given any black brane configuration.
Moreover, the fact that the underlying state-space of the brane solutions
considered in this article is everywhere regular supports the Mathur's
fuzzball proposal.

\vspace{1cm}
\begin{Large} \noindent{\bf Acknowledgement:} \end{Large}\\

The work of S.B is partially supported by the European Community Human Potential
Program under contract MRTN-CT-2004-005104 \textit{``Constituents, fundamental
forces and symmetries of the universe''}.

B.N.T. would like to thank Prof. Ashoke Sen for the discussions on
thermodynamic stability and existence of phase transitions during the
Spring School on Superstring Theory and Related Topics-2008, ICTP Trieste, Italy;
Yogesh K. Srivastava for brane fractionation's and subensemble theory during the
Indian String Meeting 2007, Harish-Chandra Research Institute, Allahabad India;
Vinod Chandra for constant support and suggestions during the preparation of
this manuscript; and acknowledges the CSIR India, for financial support under
the research grant \textit{``CSIR-SRF-9/92(343)/2004-EMR-I''}.

\end{document}